\documentclass[10pt]{article}
\usepackage[T1]{fontenc}
\usepackage{tgpagella}

\usepackage{natbib}
\usepackage{multirow}

\usepackage{a4}
\usepackage{array}
\usepackage{tabularx}
\usepackage{graphicx}
\usepackage{amsmath,amssymb,amsthm,mathtools}
\usepackage{paralist}
\usepackage{bm}
\usepackage{bbm}
\usepackage{xspace}
\usepackage{url}
\usepackage{fullpage, prettyref}
\usepackage{boxedminipage}
\usepackage{wrapfig}
\usepackage{ifthen}
\usepackage{color}
\usepackage{xcolor}
\usepackage{framed}
\usepackage[pagebackref,letterpaper=true,colorlinks=true,pdfpagemode=none,urlcolor=blue,linkcolor=blue,citecolor=purple,pdfstartview=FitH]{hyperref}
\usepackage[nameinlink]{cleveref}
\usepackage{fullpage}
\usepackage{esvect}
\usepackage{mdframed}
\usepackage{thmtools}
\usepackage{thm-restate}
\newtheorem{theorem}{Theorem}[section]

\newtheorem{lemma}[theorem]{Lemma}
\newtheorem{claim}[theorem]{Claim}

\newtheorem{definition}[theorem]{Definition}
\newtheorem{proposition}[theorem]{Proposition}

\newtheorem{remark}[theorem]{Remark}

\newcommand{\ignore}[1]{}

\newcommand{\R}{\mathbb R}

\newcommand{\RR}{\mathbb{R}}

\newcommand{\floor}[1]{\lfloor#1\rfloor}

\newcommand{\Sec}[1]{\hyperref[sec:#1]{\Cref*{sec:#1}}} %section
\newcommand{\Eqn}[1]{\hyperref[eq:#1]{(\ref*{eq:#1})}} %equation
\newcommand{\Fig}[1]{\hyperref[fig:#1]{Fig.\,\ref*{fig:#1}}} %figure
\newcommand{\Tab}[1]{\hyperref[tab:#1]{Tab.\,\ref*{tab:#1}}} %table
\newcommand{\Thm}[1]{\hyperref[thm:#1]{Theorem\,\ref*{thm:#1}}} %theorem
\newcommand{\Fact}[1]{\hyperref[fact:#1]{Fact\,\ref*{fact:#1}}} %fact
\newcommand{\Lem}[1]{\hyperref[lem:#1]{Lemma\,\ref*{lem:#1}}} %lemma
\newcommand{\Prop}[1]{\hyperref[prop:#1]{Prop.~\ref*{prop:#1}}} %property
\newcommand{\Cor}[1]{\hyperref[cor:#1]{Corollary~\ref*{cor:#1}}} %corollary
\newcommand{\Conj}[1]{\hyperref[conj:#1]{Conjecture~\ref*{conj:#1}}} %conjecture
\newcommand{\Def}[1]{\hyperref[def:#1]{Definition~\ref*{def:#1}}} %definition
\newcommand{\Alg}[1]{\hyperref[alg:#1]{Alg.~\ref*{alg:#1}}} %algorithm
\newcommand{\Obs}[1]{\hyperref[obs:#1]{Obs.~\ref*{obs:#1}}} %observation
\newcommand{\Ex}[1]{\hyperref[ex:#1]{Ex.~\ref*{ex:#1}}} %example
\newcommand{\Clm}[1]{\hyperref[clm:#1]{Claim~\ref*{clm:#1}}} %example
\newcommand{\Step}[1]{\hyperref[step:#1]{Step~\ref*{step:#1}}} %example

\newenvironment{proofof}[1]{\noindent \textbf{Proof of #1.}}{\qed}

\usepackage{booktabs}

\usepackage{tikz}
\usetikzlibrary{arrows.meta}
\usetikzlibrary{shapes}

\renewcommand{\star}{\ast}

\usepackage[utf8]{inputenc} % allow utf-8 input
\usepackage[T1]{fontenc}    % use 8-bit T1 fonts
\usepackage{hyperref}       % hyperlinks
\usepackage{url}            % simple URL typesetting
\usepackage{booktabs}       % professional-quality tables
\usepackage{amsfonts}       % blackboard math symbols
\usepackage{nicefrac}       % compact symbols for 1/2, etc.
\usepackage{microtype}      % microtypography
\usepackage{xcolor}         % colors
\usepackage[ruled,linesnumbered]{algorithm2e}
\SetKwComment{Comment}{/* }{ */}
\usepackage{wrapfig}
\usepackage{algorithmic}
\usepackage{float}

\usepackage{subcaption}

\hypersetup{linktocpage = true, linkcolor = blue}

\DeclareMathOperator*{\argmax}{arg\,max}
\DeclareMathOperator*{\argmin}{arg\,min}

\title{Query Complexity of Learning Axis-Aligned Partitions\thanks{This work was supported by NSF TRIPODS Institute grant 2217058 (EnCORE).}}

\author{
Hadley Black\thanks{Baruch College, CUNY. {\tt hadley.black@baruch.cuny.edu}.} \\  
\and 
Arya Mazumdar\thanks{UC San Diego. {\tt arya@ucsd.edu}.}
\and
Barna Saha\thanks{UC San Diego. {\tt bsaha@ucsd.edu}.}
\and 
Geelon So\thanks{ETH Z\"urich. {\tt geelon.so@inf.ethz.ch}.} \\
}

\begin{document}

\maketitle

\begin{abstract}
    This work establishes the query complexity of learning high-dimensional boxes in the \emph{exact query learning} setting. Let $X \subset \mathbb{R}^d$ be a set of $n$ points, labeled by a collection of $k$ boxes: the label of each point in $X$ is determined by the unique box containing it. A learner aims to exactly recover the labels for every point in $X$ given access to a label oracle. We consider three increasingly-structured settings. In the most general setting, there are no conditions on the boxes, and they may overlap. 
    We show that CAL (Cohn-Atlas-Ladner 1994) a classical active learning algorithm, is minimax-optimal with query complexity $\Theta(kd \log (n/kd))$. When the boxes are disjoint, we examine two regimes. In the large-$k$ regime, CAL remains minimax-optimal. For the constant-$k$ regime, an optimized variant of the Halving algorithm (Heged{\"{u}}s 1995) can beat CAL by a $\log(d)$ factor. Finally, in the most-structured settings, the boxes correspond to the leaves of a constant-sized decision tree. We prove a tight query complexity of $\Theta(d + \log n)$ that is additive in $d$ and $\log n$, generalizing a recent result (Black-Larsen-Mazumdar-Saha-So 2026) that establishes the query complexity of learning decision stumps.
\end{abstract} 

\newpage

\tableofcontents

\newpage

\section{Introduction}

This work establishes the query complexity of learning high-dimensional boxes under the setting of \textit{pool-based active learning} with \textit{label queries} \citep{DBLP:journals/ml/Angluin87,DBLP:conf/nips/Dasgupta04,settles2009active}. Let $X \subset \mathbb{R}^d$ be any finite set of $n$ points and $Y := [k]$ be a set of $k$ labels. The goal of a learner is to recover an unknown labeling $f : X \to Y$ when provided access to a label oracle: upon being given a query $x \in X$, the oracle supplies the label $f(x)$. The basic problem is to understand how many queries the learner needs before being able to deduce the label of every point in $X$.

We study the family of \emph{axis-aligned partitions}, or more plainly, \emph{box partitions}. A labeling $f$ is a box partition if it is induced by $k$ boxes: for each label $y \in Y$, there is a box $B_y \subset \mathbb{R}^d$ such that the points in the $y$th class are given by $f^{-1}(y) = X \cap B_y$. There are $a_{y,i} \leq b_{y,i} \in \mathbb{R} \cup \{\pm \infty\}$ so that
\[\hspace{2em}\forall x \in X, \quad f(x) = y \quad \Longleftrightarrow\quad x \in B_y, \qquad \textrm{where }B_y = \prod_{i=1}^d [a_{y,i}, b_{y,i}].\]

A simple example of box partitions is the family of \emph{axis-aligned halfspaces} (decision stumps). These are box partitions that split $X$ in two, based on which side of an axis-aligned halfspace points fall; the splitting halfspace can also be seen as the face along which two adjacent, unbounded boxes meet. Another example is the family of box partitions induced by the leaves of \emph{decision trees}, which generalize decision stumps via recursion. Decision trees assign two instances the same label if and only if they are contained in the same leaf of the tree. Beyond decision stumps and decision trees, this work considers a hierarchy of box partitions (see \Cref{fig:main_figure}). With increasing structure, they are:
\begin{itemize}
    \item \textbf{General box partitions:} These constitute the broadest class of axis-aligned partitions. The only requirement is that the induced labeling is well-defined, meaning that each point in $X$ is contained in precisely one of the underlying boxes. The boxes themselves may overlap in $\mathbb{R}^d$.
    \item \textbf{Disjoint box partitions:} The boxes must be disjoint, $B_{y} \cap B_{y'} = \emptyset$ for distinct $y, y' \in Y$.
    \item \textbf{Decision trees:} The boxes are disjoint and admit a decision-tree structure. This means that there is a binary tree where each leaf is identified with a label in $Y$. For each internal node in the binary tree, there is a dimension $i \in [d]$ and a threshold $\tau \in \mathbb{R}$ that splits its left and right descendants: a descendant $y \in Y$ is to the left if and only if $f^{-1}(y) \subset \{x \in \mathbb{R}^d : x_i < \tau\}$.
\end{itemize}

\subsection{Main Results}

We study the query complexity of learning axis-aligned partitions through a series of classic active learning algorithms. Besides giving specialized upper bounds, the main contribution of this work establishes lower bounds across these settings. This illuminates the way in which different sources of geometric structure affect the hardness of learning, and how different algorithms can leverage it.

The first result is \Cref{thm:overlap} for learning general class of box partitions. It establishes a tight query complexity of $\Theta(kd \log (n/kd))$, where $n := |X|$ is the number of points, $k := |Y|$ is the number of labels, and $d$ is the number of dimensions. Formally:

\begin{figure}[t]
    \centering

    % First panel
    \begin{minipage}[b]{0.27\textwidth}
        \centering
        \includegraphics[width=\textwidth]{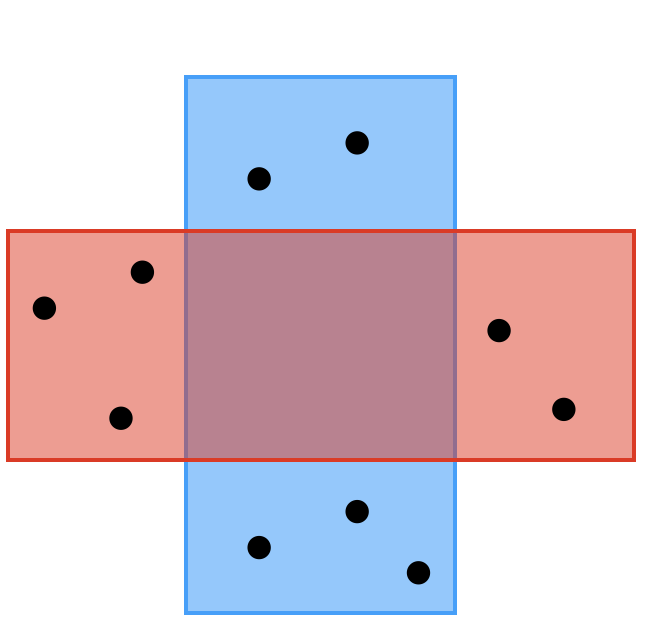}

        \smallskip
        \footnotesize
        (a)
    \end{minipage}
    \hfill
    %
    % Second panel
    \begin{minipage}[b]{0.30\textwidth}
        \centering
        \includegraphics[width=\textwidth]{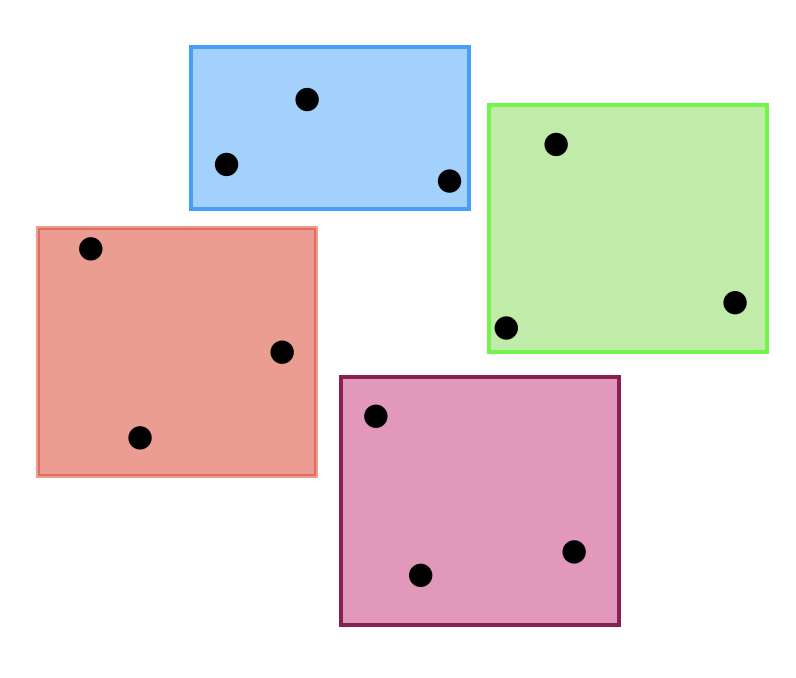}

        \smallskip
        \footnotesize
        (b)
    \end{minipage}
    \hfill
    %
    % Third panel
    \begin{minipage}[b]{0.36\textwidth}
        \centering
        \includegraphics[width=\textwidth]{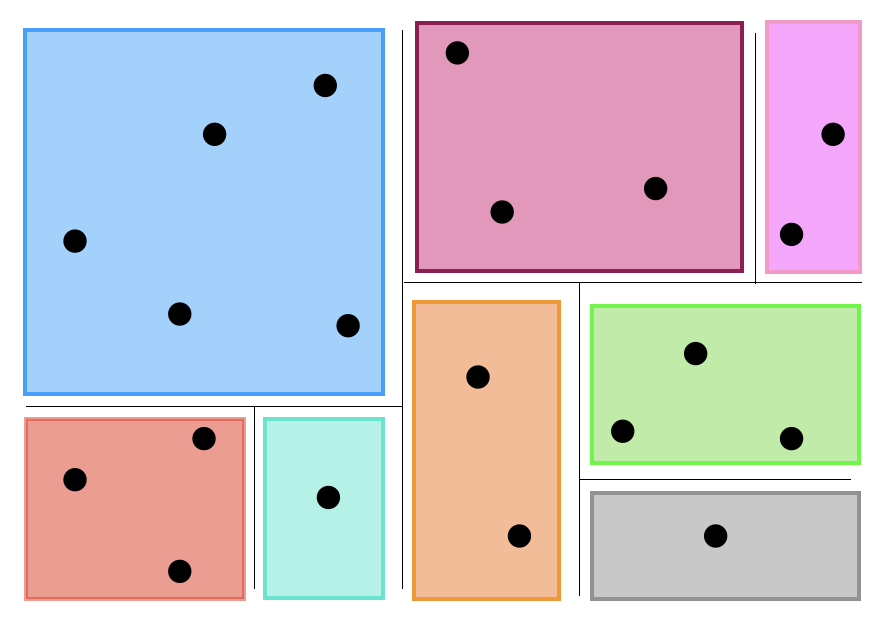}

        \smallskip
        \footnotesize
        (c)
    \end{minipage}%

    \caption{
        Axis-aligned partitions induced by (a) general boxes, (b) disjoint boxes, and (c) the leaves of a decision tree. These classes form a strictly descending chain: (a) is a box partition with two labels that cannot be induced by any disjoint box partition, and (b) is a disjoint box partition with four labels that cannot be induced by any decision tree.
    }
    \label{fig:main_figure}
\end{figure}

\begin{theorem} [General Boxes] \label{thm:overlap} Let $n \geq k \geq 2$. Consider the problem of learning a general $k$-box partition over a worst-case set of $n$ points in $d$-dimensions.

\begin{itemize}
    \item The expected query complexity of this problem is $\Theta\left(kd\log(1+\frac{n}{kd})\right)$. 
    \item For any $1 \leq r \leq \log_2(n/kd)$, the $r$-round query complexity is $\Theta\left(rkd (1+\frac{n}{kd})^{1/r}\right)$.
\end{itemize}
In particular, there is an algorithm achieving the optimal query complexity with a $1 + \log_2(1+n/dk)$ round budget. Our lower bounds hold even against randomized algorithms. 
\end{theorem}

A canonical active learning algorithm achieves minimax-optimality: it is enough to run CAL, which is named after the authors who introduced it in \cite{DBLP:journals/ml/CohnAL94}. This algorithm sequentially chooses a random point whose label the learner cannot determine yet, and asks the oracle for the ground-truth label. Furthermore, batched-CAL is minimax-optimal even in settings where the learner is limited to $r$ rounds of interactions with the oracle. We prove this theorem in \Cref{sec:general-box-partitions}.

The upper bounds achieved by CAL and its round-aware variant certainly apply to any class of box partitions, including disjoint box partitions and decision trees. The same is not necessarily true for the lower bound. If there is additional structure, can it be leveraged to speed up learning?

The next result \Cref{thm:disjoint} focuses on disjoint box partitions in two main regimes. In the large-$k$ regime with $k \geq 2^d$ boxes, CAL remains minimax-optimal: the query complexity is lower bounded by $\Omega(kd \log (n/kd))$, which matches the upper bound for general box partitions. On the other hand, in the small-$k$ regime where the number of boxes $k = O(1)$ is constant, the upper bound for disjoint boxes can be a $\log(d)$-factor smaller than the lower bound for general boxes:

\begin{theorem} [Disjoint Boxes] \label{thm:disjoint}
    Let $d,k\geq 2$. There is a randomized algorithm that learns any disjoint $k$-box partition over a worst-case set of $n$ points in $d$-dimensions using in expectation
    \[O\left(\min\left\{n, k d\log \left(1+\frac{n}{kd}\right), \frac{k^3 d \log n}{\log (kd)}\right\}\right) \textrm{ queries.}\]
    When $k \geq 2^{\Omega(d)}$, the upper bound is matched with a lower bound  of $\Omega\left(kd \log \frac{n}{kd}\right)$ queries.
\end{theorem}
\begin{table}[t]
\centering
\renewcommand{\arraystretch}{1.15}
\begin{tabular}{@{}llll@{}}
\toprule
\textbf{Partition class}
    & \textbf{Regime}
    & \textbf{Bound}
    & \textbf{Result} \\
\midrule

Boxes
    & all
    & $\Theta\left(kd \log \frac{n}{kd}\right)$
    & \Cref{thm:overlap} \\

\addlinespace[2pt]
\midrule
\addlinespace[2pt]

\multirow[c]{2}{*}[-1ex]{Disjoint boxes}
    & $k \geq 2^d$
    & $\Theta\left(kd \log \frac{n}{kd}\right)\vphantom{\left(\frac{d \log n}{\log d}\right)}$
    & \multirow[c]{2}{*}[-1ex]{\Cref{thm:disjoint}} \\

\cmidrule(lr){2-3}

    & $k = O(1)$
    & $O\left(\frac{d \log n}{\log d}\right)$
    & \\

\addlinespace[2pt]
\midrule
\addlinespace[2pt]

Decision-tree leaves
    & $k = O(1)$
    & $\Theta\left(d + \log n \vphantom{\frac{n}{kd}}\right)$
    & \Cref{thm:dt} \\

\addlinespace[2pt]
\midrule
\addlinespace[2pt]

Decision stumps
    & $k = 2$
    & $\Theta\left(d + \log n\vphantom{\frac{n}{kd}}\right)$
    & \cite{DBLP:conf/colt/BlackLMSS26} \\

\bottomrule
\end{tabular}
\caption{Summary of the query complexity of learning box partitions.}
\label{tab:query-complexity}
\end{table}

We prove this theorem in \Cref{sec:disjoint-box-partitions}. In the small-$k$ regime, the algorithm that achieves the faster rate is another classic active learning algorithm called the Halving algorithm, or rather, a slightly more optimized version of it found in \cite{moshkov1982conditional}
and \cite{DBLP:conf/colt/Hegedus95}. Let's call it Halving++. While CAL is essentially content to query a random point with an ambiguous label, Halving++ seeks particularly informative points to query. One combinatorial quantity that controls the query complexity of Halving++ is the \emph{extended teaching dimension} \citep{DBLP:conf/colt/Hegedus95}. We prove a query complexity upper bound by way of bounding the extended teaching dimension of disjoint boxes. The proofs are in \Cref{sec:disjoint-box-partitions}. Compared to general boxes, the faster rate for disjoint boxes cannot only be attributed to the presence of additional structure; the learner also needs to take advantage of it. \Cref{sec:separation} gives an instance of a disjoint-box problem where CAL provably fails to exploit the structure that Halving++ successfully leverages. 

The final result is \Cref{thm:dt}, which considers box partitions induced by the leaves of a decision tree. Importantly, these decision trees are distinct from the more common \emph{binary-valued} decision trees; the induced labeling here is a multi-class function, where the number of classes $k$ is equal to the number of non-empty leaves.\footnote{Binary-valued decision trees have only two classes. A single label class can be a union of multiple leaves/boxes. It is easy to construct binary decision trees where the query complexity is $\Omega(n)$, see also \cite{DBLP:conf/colt/BlackLMSS26}.} For constant-sized trees, the query complexity has \textit{additive} upper and lower bounds of $\Theta_k(d + \log n)$. Thus, there is a further separation between this setting and the general box setting, which has a \textit{multiplicative} lower bound of $\Omega_k(d \log n)$. We show:

\begin{theorem} [Decision Trees] \label{thm:dt}
    Let $k,d\geq 2$ and $n \geq d(k-1)$. There is a deterministic algorithm that learns any box-partition induced by the leaves of a $k$-leaf decision tree using at most
    \[\hspace{1em} 2^{O(k)} \left(d + \log n\right) \textrm{ queries}.\]
    If the decision tree is a decision list, then the query complexity is upper bounded by $O(k^2(d + \log n))$, and it is lower bounded by $\Omega(k(d + \log(n/k)))$ when $n \geq d$. 
\end{theorem}

To achieve this additive upper bound on query complexity, we make use of a `parallel binary search' algorithm, which was introduced by \cite{DBLP:conf/colt/BlackLMSS26}. The parallel binary search algorithm is a minimax-optimal learner for decision stumps. \Cref{thm:dt} can be viewed as a generalization of that result, establishing the query complexity not just for decision stumps, but for constant-sized decision trees. The lower bound we provide here holds even for \emph{decision lists}, which are decision trees whose internal nodes form a path graph. The proofs of these results are provided in \Cref{sec:decision-tree}.

A high-level summary of our main results is given in \Cref{tab:query-complexity}. The next section also provides some high-level sketch of the technical ideas in this work. 

\subsection{Technical Overview}

\paragraph{General box partitions.} \Cref{thm:overlap} provides a query complexity upper bound for CAL and a round-constrained variant. It also gives a matching lower bound construction. The upper bounds rely on a simple observation that applies to any box partition. Suppose that a learner has seen a set of points $S = \{x_1,\ldots, x_m\}$ with the same label $y$. Then, every point in $X$ contained in the bounding box of $S$ must also be labeled $y$, and CAL will not query these points. The upper bounds follow from determining the rate that these labeled bounding boxes grow before they fully cover $X$. 

The lower bounds throughout this work are built on the fact that learning a monotone function
$f : [n] \to \{0,1\}$ from label queries requires $\Theta(\log n)$ queries. The constructions find ways to embed many independent copies of these one-dimensional threshold learning problems into the various box learning settings. For general box partitions, we embed $\Theta(kd)$ threshold problems, each consisting of $\Theta(n/kd)$ points. Thus, we obtain an $\Omega(kd \log(n/kd))$ lower bound.

These lower bounds are rather non-trivial to find. Consider even $k = 2$ boxes. It is clear how \textit{one} threshold learning problem over $n$ points can be embedded into a 2-box learning problem: if we arrange the points in a line, then each threshold can be induced by a pair of adjacent boxes. But this construction only gives an $\Omega(\log n)$ lower bound, not the stronger $\Omega(d \log(n/d))$. The problem is that only a single face of each of the boxes contributes to hardness. Intuitively, naive construction under-utilizes the set of $2d$ faces available from each box. The question is how to place the points so that all $\Theta(d)$ faces of the two boxes can interact with each other to admit $\Theta(d)$ independent threshold learning problems. The construction we give achieves this using overlapping boxes. It turns out that to extract this hardness for general $k$, it is necessary that boxes are able to overlap.

\paragraph{Disjoint box partitions.} \Cref{thm:disjoint} consists of two parts. It shows a lower bound for disjoint boxes in the large-$k$ regime that continues to match the general-box upper bound. It also shows that in the small-$k$ regime with disjoint boxes, it is possible to beat the general-box lower bound.

The lower bound in the large-$k$ regime is a construction that embeds $\Theta(kd)$ independent threshold learning problems into a disjoint box learning problem. When $k = 2^d$, we will embed a threshold learning problem onto every edge of a Boolean hypercube (note there are $d 2^{d-1} = \Theta(kd)$ edges). To do this, place points along each edge of the hypercube, and extend a box out of each vertex onto its incident edges; see \Cref{fig:LB-disjoint}. The problem is that the threshold learning problems induced this way are not automatically independent, as the boxes may intersect near the center of the Boolean hypercube. However, we carefully design point configurations that manage to preserve independence, thus obtaining an $\Omega(kd \log (n/kd))$ lower bound.

The second part shows that Halving++ beats the general-box lower bound by a $\log(d)$-factor in the $k = \Theta(1)$ constant regime. This follows from an upper bound on the query complexity of Halving++ \citep{moshkov1982conditional,DBLP:conf/colt/Hegedus95}. To specialize the bound, we bound the extended teaching dimension of disjoint-box partitions, as well as the number of such partitions of $X$. Beating the general-box lower bound in certain regimes is possible because disjoint boxes have additional structure: every pair of box can be separated by an axis-aligned halfspace. The labeling induced by disjoint boxes can be specified either through the set of $2kd$ thresholds associated to the faces of the boxes, or through a set of $\Theta(k^2)$ halfspaces that separate every pair of label class. In a sense, CAL learns only by recovering the $2kd$ thresholds. Halving++ can be strictly more label-efficient, and it does so intuitively by leveraging this alternative specification, which is independent of $d$. For constant order $k$, we show that this is ultimately the source of the $\log(d)$-factor savings.

\paragraph{Decision trees.} In \Cref{thm:dt} we consider the class of disjoint box partitions that are induced by $k$-leaf decision trees. While the upper bounds from \Cref{thm:overlap} and \Cref{thm:disjoint} extend to this setting, we show that the additional structure of decision trees enables improved query bounds, at least when $k$ is relatively small. \cite{DBLP:conf/colt/BlackLMSS26} gave a $O(d + \log n)$ query learner for the class of $2$-leaf decision trees (or decision stumps). Intuitively, if we compare $2$-box partitions, the gap between the $\Theta(d \log (n/d))$ sample complexity for general boxes and $\Theta(d + \log n)$ for decision stumps comes from the fact that a $2$-leaf tree is specified by a single axis-aligned hyperplane. Our lower bound for general boxes shows that one essentially needs to learn all the $\Omega(d)$ faces of the two boxes. 

More generally, for $k\geq 2$, when a labeling $f \colon X \to Y$ is induced by a decision tree, we are guaranteed the existence of a single axis-aligned hyperplane that separates the labels into two non-empty sets (see \Cref{def:decision}), $A \subset Y$ and $Y \setminus A$, and this same structure repeats on each side. If this subset $A$ is known in advance, then one can apply the \emph{decision stump} learner to recover $g_A(x) = \mathbf{1}(f(x) \in A)$. Extending this observation, if the label-splits are provided at each level of the tree, this would yield a $O(k(d+\log n))$ query algorithm. This shows that naively guessing every such tree yields a $k^{O(k)}(d+ \log n)$ query algorithm (see \Cref{rem:DT-simple}). In \Cref{thm:dt}, we reduce the $k$-dependence to $2^{O(k)}$ with a more careful recursive strategy that avoids brute-force enumeration. At a very high level, we obtain this improvement by certifying correct label-splits at the leaves and propagating this information through recursion. This allows us to make permanent progress without guessing the correct tree all at once.

\subsection{Open Questions}

The aim of our work is to obtain a comprehensive understanding of the query complexity of learning natural axis-aligned partitions over worst-case point-sets. The following gaps remain, which we leave as interesting directions for follow up work, and which we believe will require new insights. 

\begin{itemize}
    \item \emph{Improved bounds for disjoint boxes in the intermediate-$k$ regime}. When $k \geq 2^{\Omega(d)}$, we obtain a $\Omega(kd\log(n/kd))$ lower bound, matching our upper bound for general $k$-box partitions. However, when $k < 2^{o(d)}$, our bounds are not tight, and our upper bounds in \Cref{thm:disjoint} demonstrate a separation from general boxes when $k$ is sufficiently small. Therefore, as $k$ becomes small, there is a moment where box disjointness supplies more geometric structure that can be exploited by a learner. Obtaining sharper bounds in this regime would require finding new ways to use this geometry, and we believe this is a worthwhile future direction. 
    \item \emph{Improved bounds for decision trees.} Decision trees are an extremely natural specialized class of box partitions to consider. While we obtain bounds that are tight for any constant number of classes, it is not clear what the true dependence on $k$ is for this problem. We believe this is a very interesting open question to resolve.
\end{itemize}

\subsection{Related Work}

There is an extensive line of research in exact query learning and active learning; as an inexhaustive list, see \cite{DBLP:journals/ml/Angluin87,DBLP:conf/nips/Dasgupta04,DBLP:conf/colt/BalcanBZ07,settles2009active,DBLP:conf/colt/BalcanL13,hanneke2015minimax}, and references therein. This work belongs to a subarea concerned with learning geometric concept classes, in which the label function is induced by fundamental objects like balls, boxes, halfspaces, convex sets, and decision trees \citep{ben1997composition,bshouty1998exact,bshouty1998noise,gonen2013efficient,ashtiani2016clustering,DBLP:journals/jmlr/0002CLP24,DBLP:conf/colt/BlackLMSS26,soltani2026active}. 

This work establishes the query complexity of active learning axis-aligned partitions under various structural assumptions, and our main contributions are our lower bound constructions. Our results complement existing work. A prior result for learning general $k$-box partitions is given by \cite{DBLP:journals/jmlr/0002CLP24}. They provide a generic algorithm for multi-class active learning, which uses $O(kd \log k \log n)$ queries when specialized to boxes. In a more specific two-class setting, \cite{DBLP:conf/colt/BlackLMSS26} establishes that the query complexity for axis-aligned halfspaces is $\Theta(d + \log n)$. 

This work shows that CAL \citep{DBLP:journals/ml/CohnAL94}, a standard active learning algorithm, is minimax-optimal for general boxes. The first generic upper bound of the sample complexity of CAL was given by \citep{DBLP:journals/jcss/BalcanBL09}, and they also provide a specialized analysis for the 1-dimensional linear threshold. Our upper bound can be seen as a generalization of the analysis for 1-dimensional thresholds: a set of $k$ boxes in $d$ dimensions can be specified by a set of $kd$ thresholds. We also show that a variant of the Halving algorithm \citep{DBLP:conf/colt/Hegedus95} can exploit additional structure, and that it strictly beats CAL in certain regimes for learning disjoint boxes. The most structured concept class we consider generalizes decision stumps; they are induced by the leaves of decision trees \citep{DBLP:conf/icml/MoshkovitzDRF20,hopkins2021bounded}. For these tree-based partitions, we also give a minimax-optimal learner for constant number of classes.

The lower bounds we prove arise from embedding many independent copies of a threshold learning problem into various box-learning problems. In the round-constrained setting, lower bounds also arise from corresponding bounds for threshold learning provided by \cite{thresholds-LB-ref}.

\paragraph{Notation.} For any $a \leq b \in \mathbb{R} \cup \{\pm \infty\}$, we let $[a,b]$ denote the closed interval from $a$ to $b$. Given a finite set of points $S \subset \mathbb{R}^d$, denote its \emph{bounding box} by
\[\mathsf{bbox}(S) = \prod_{i=1}^d \left[\min_{x \in S} x_i, \max_{x \in S} x_i\right].\]
\section{General Box Partitions} \label{sec:general-box-partitions}

In this section, we prove \Cref{thm:overlap}, which considers learning a general box partition: there is a set of $k$ potentially-overlapping boxes $B_1, \ldots, B_k \subset \mathbb{R}^d$ that induce a partition of $X$
\[X = (X \cap B_1) \sqcup \dotsm \sqcup (X \cap B_k).\]
We divide up the results. The upper bound without round constraints is given by \Cref{thm:UB-adaptive}. The upper bound with round constraints is in \Cref{thm:r-round}. The lower bound is given in \Cref{thm:general-LB}.

\subsection{Learning General Box Partitions} \label{sec:box-partition-UB-adaptive}

A standard active learning algorithm CAL \citep{DBLP:journals/ml/CohnAL94} achieves the query complexity upper bound. It is an iterative algorithm that maintains a \emph{disagreement region}. This region is the subset of points of $X$ whose labels cannot be deduced yet: it contains a point $x$ if there are at least two valid labelings that disagree on how to label $x$, but that are nevertheless consistent with the observed data. Every round, the learner randomly selects a point from this region and asks the oracle for its label. In the case of learning boxes, if $S$ is a set of points that have been labeled $y$, then the learner may deduce that every point in the bounding box $\mathsf{bbox}(S)$ is also labeled $y$. The disagreement region shrinks over the course of the algorithm, and the learner stops when this region vanishes. The following upper bound is similar in spirit to Theorem 4.1 of \citep{DBLP:journals/jcss/BalcanBL09}.

\begin{theorem} \label{thm:UB-adaptive} There is a randomized algorithm that learns an arbitrary $k$-box partition in $d$ dimensions over $n$ points using at most $4kd \log (1 + \frac{n}{kd})$ label queries in expectation.
\end{theorem}

\begin{proof} The algorithm is defined as follows.
\begin{enumerate}
    \item Initialize a partial function $h \colon X \to [k] \cup \{\ast\}$ as $h(x) = \ast$ for all $x \in X$. For each $y \in [k]$, let $X_{h,y} = \{x \in X \colon h(x) = y\}$ throughout the execution. We will maintain the invariant \[X_{h,y} \subseteq X_y.\]
    \item Choose a permutation $\pi \colon [n] \to X$ uniformly at random.
    \item For each $i \leq n$, let $x = \pi(i)$ and do the following:
    \begin{enumerate}
        \item If $h(x) \neq \ast$, then continue.
        \item If $h(x) = \ast$, then query $f(x)$. Set $h(x) = f(x)$, and also set $h(z) = f(x)$ for all points $z \in \mathsf{bbox}(X_{h,f(x)})$.
    \end{enumerate}
\end{enumerate}

The correctness of the algorithm is clear by inspection. We now analyze the expected number of queries over the randomness of $\pi$. Whether or not a point with label $y$ is queried depends only on the ordering that $\pi$ induces on $X_y$. Let $\pi_y \colon [n_y] \to X_y$ denoted this restriction of $\pi$ where $n_y := |X_y|$. Observe that for each $t \leq n_y$, $\pi_y(t)$ is queried if and only if it is \emph{not contained} in the bounding box of $\pi_y(1),\ldots,\pi_y(t-1)$. Equivalently, $\pi_y(t)$ is queried if and only if it is a \emph{maximal point}\footnote{We say $x \in S$ is maximal among the points in $S$ if $x \notin \mathsf{bbox}(S\setminus x)$. In other words, for some $i \in [d]$, $x_i$ is the unique maximum or minimum $i$-coordinate value among points in $S$.} among $\pi_y(1),\ldots,\pi_y(t)$. Since $d$-dimensional boxes have at most $2d$ maximal points, and $\pi_y(t)$ is uniform random among $\pi_y(1),\ldots,\pi_y(t)$, the probability $\pi_y(t)$ is queried is at most $2d/t$. Let $Q(\pi) = \sum_{y} Q_y(\pi)$ denote the total number of queries made by the algorithm on permutation $\pi$, where $Q_y(\pi)$ is the number of queries in $X_y$. By linearity of expectation,
\begin{align*}
    \mathbb{E}_{\pi_y}\left[Q_y(\pi)\right] &= \sum_{t=1}^{n_y} \Pr_{\pi_y}\left[\pi_y(t) \text{ is queried}\right] \\
    &\leq \sum_{t=1}^{n_y} \min\left(1,\frac{2d}{t}\right) \leq \sum_{t=1}^{n_y}\frac{4d}{2d+t} \leq 4d\int_0^{n_y}\frac{dx}{2d+x} = 4d\ln\left(1+\frac{n_y}{2d}\right).
\end{align*}

Applying linearity of expectation over all $k$ labels yields
\begin{align*}
    \mathbb{E}_{\pi}\left[Q(\pi)\right] \leq 4d \sum_{y=1}^k \ln\left(1+\frac{n_y}{2d}\right) \leq 4kd \ln\left(1+\frac{n}{2kd} \right)
\end{align*}
by Jensen's inequality. \end{proof}

Next we consider an $r$-round variant of CAL. The proof is in \Cref{sec:low-round-alg}.

\begin{theorem} [$r$-round algorithm] \label{thm:r-round} For any $1 \leq r \leq \log(n/2dk)$, there is an $r$-round randomized algorithm which learns an arbitrary $k$-box partition in $d$-dimensions over $n$ points using 
\[
2rk d \cdot \left(\frac{n}{2dk}\right)^{1/r}
\]
queries in expectation. In particular, when $r=\log(n/2dk)$, this recovers the optimal query complexity $O(kd\log(n/kd))$, while improving on the round complexity in \Cref{thm:UB-adaptive} by a $\Theta(kd)$ factor.
\end{theorem}

\subsection{Matching Lower Bound} \label{sec:LB-overlapping}

The simplest, nontrivial setting for learning box partitions is when $k = 2$ and $d = 1$. Here, the problem is simply to learn a threshold between two classes. In this case, we can of course run binary search, which optimally solves the problem using at most $\log_2 n$ queries. The following result provides a lower bound for the threshold learning problem, with and without round constraints.

\begin{theorem} [Threshold Learning Lower Bound, Theorem 3 of \cite{thresholds-LB-ref}] \label{thm:LB-threshold} Consider exactly learning a monotone function $f \colon [n] \to \{0,1\}$ with query access. Suppose that a randomized algorithm $\mathcal{A}$ solves this problem on every input $f$ with success probability at least $2/3$. Then, there exists $f$ such that the expected number of queries made is at least $\Omega(\log n)$. Moreover, if $\mathcal{A}$ uses $1 \leq r \leq \log_2 n$ rounds of adaptivity, then there exists $f$ such that the expected number of queries made is at least $\Omega(r n^{1/r})$.
\end{theorem}

To obtain query lower bounds for general $k ,d \geq 2$, we will show that a set of $\Theta(kd)$ independent threshold learning problems with $m$ points each can be reduced to learning a general $k$-box partition over a set of points of size $n = \Theta(kdm)$. We obtain a lower bound $\Omega(kd \log (n/kd))$ for the round-unconstrained setting, and a lower bound of $\Omega(r kd(n/kd)^{1/r})$ for the $r$-round setting:

\begin{theorem}[Lower bound for general boxes]
    \label{thm:general-LB}
    Let $k, d \geq 2$ be even. Let $2 \leq m \leq \frac{4n}{kd}$ be a natural number. Suppose a randomized algorithm learns any $k$-box partition over $n$ points in $\mathbb{R}^d$ with success probability $2/3$. Then, there exists a problem instance such that the expected number of queries made is at least $\Omega(kd \log m)$. Moreover, if the algorithm uses $1 \leq r \leq \log_2 m$ rounds of adaptivity, then there exists a problem instance such that the expected number of queries made is at least $\Omega(rkd m^{1/r})$.
\end{theorem}

\begin{figure}[t]
    \centering
    \begin{tikzpicture}[
  >=Latex, line cap=round, line join=round, scale=0.4]

% Parameters: integer side length and a threshold between samples.
\def\PlotM{10}
\def\PlotThreshold{4.5}

\colorlet{boxzero}{blue!20}
\colorlet{boxone}{orange!25}
\colorlet{ptzero}{blue!80!black}
\colorlet{ptone}{orange!85!black}

% Overlapping rectangles.
\fill[boxzero]
  (0,0) rectangle (\PlotM,\PlotThreshold);
\fill[boxone,fill opacity=0.7]
  (0,0) rectangle ({\PlotM-\PlotThreshold},\PlotM);

\draw[blue!50!black,thin]
  (0,0) rectangle (\PlotM,\PlotThreshold);
\draw[orange!60!black,thin]
  (0,0) rectangle ({\PlotM-\PlotThreshold},\PlotM);

% Axes, without labels or ticks.
% \draw[->,thick] (0,0) -- ({\PlotM+1.1},0);
% \draw[->,thick] (0,0) -- (0,{\PlotM+1.1});

% Embedded points (m-t,t).
\foreach \t in {1,...,\PlotM} {
  \ifdim\t pt<\PlotThreshold pt\relax
    \fill[ptzero] ({\PlotM-\t},\t) circle[radius=4.5pt];
  \else
    \fill[ptone] ({\PlotM-\t},\t) circle[radius=4.5pt];
  \fi
}

\fill[black] (0,0) circle[radius=4.5pt];

% Threshold marker
\node[
  diamond,draw=black,fill=white,
  aspect=1.2,minimum size=5.5pt,inner sep=0pt
] at ({\PlotM-\PlotThreshold},\PlotThreshold) {};

\end{tikzpicture}
    \caption{A visualization of the construction in \Cref{thm:general-LB}. This shows a projection of the 2-box construction onto the $(2j-1, 2j)$th coordinates. It embeds a threshold problem, where the threshold is marked by a white diamond. Many other points project to the black origin.}
    \label{fig:LB-overlap}
\end{figure}
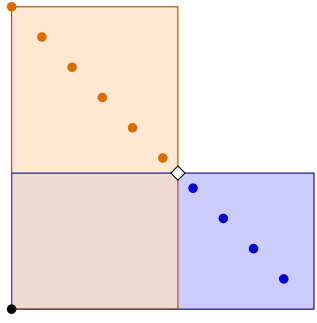

\begin{proof}
    Let $C = \frac{kd}{4}$. Let $f_{i,j} : [m] \to \{0,1\}$ be a collection of $C$ independent monotone functions, where $(i,j)$ ranges over $[k/2] \times [d/2]$. We construct a reduction from learning these thresholds to learning a $k$-box partition over $Cm$ points as follows.

    Let's start with the $k = 2$ case. We drop the index $i$ as it now ranges over the singleton $\{1\}$. Define the map $\phi : [d/2] \times [m] \to \mathbb{R}^d$ by
    \[\phi(j,t) \equiv \phi_j(t) = \Big(0,\ldots, 0, \underbrace{\vphantom{\bigg|}m - t, \,t}_{\substack{\textrm{$(2j -1, 2j)$th}\\\textrm{coordinate}}}, 0,\ldots, 0 \Big).\]
    We can think of $\phi_j$'s as a collection of $d/2$ affine embeddings of $[m]$ into $\mathbb{R}^d$. Define the set
    \[X = \bigcup_{j \in [d/2]} \phi_j([m]).\]
    By construction, the map $\phi : [d/2] \times [m] \to X$ is bijective. Define the labeling $f : X \to \{0,1\}$ by
    \[x = \phi(j,t) \quad \implies \quad f(x) = f_j(t).\]
    This is visualized in \Cref{fig:LB-overlap}. We claim the following:

    \begin{claim}
        The labeling $f$ is a 2-box partition. Moreover, the following hold:
        \begin{enumerate}
        \item[(a)] The set of $C$ threshold functions $f_j$'s can be computed from $f$.
        \item[(b)] Every query to a box-label oracle can be simulated by a query to one threshold-label oracle.
    \end{enumerate}
    \end{claim}

    In particular, this claim shows that any box-learning algorithm that learns $f$ with success probability $p$ can be converted into an algorithm that learns the original set of thresholds with the same success probability $p$ using the same number of queries. By \Cref{thm:LB-threshold}, we immediately have that the number of queries that the box-learning algorithm makes to a label oracle is at least
    \[C \cdot \Omega\left(\max\left\{r m^{1/r},  \log_2 m \right\}\right).\]
    We now prove the claim. The subclaims (a) and (b) are direct, and they are conceptually converses of each other. We can compute $f_j$ from $f$ since $f_j(t)$ is equal to $f(\phi_j(t))$. On the other hand, a query $x = \phi(j,t)$ to the box-label oracle can be simulated by querying $f_j(t)$. 
    
    \paragraph{Proof that $f$ is a 2-box partition.} Let $0 < \tau_j < m+1$ be a threshold for $f_j$, so that
    \[f_j(t) = \begin{cases}
        0 & t < \tau_j\\
        1 & t > \tau_j.
    \end{cases}\]
    Define the boxes $B_0$ and $B_1$ as follows. Let $\Pi_j : \mathbb{R}^d \to \mathbb{R}^2$ be the projection onto the $(2j-1)$th and $(2j)$th components for $j \in [d/2]$. Then:
    \[\Pi_j (B_0) = [0, m] \times [0, \tau_j] \qquad \textrm{and}\qquad \Pi_j(B_1) = [0,m-\tau_j] \times [0,m].\]
    These two boxes are defined so that $y = f_j(t)$ if and only if $\phi(j,t) \in B_y$. Indeed for all $t \in [m]$,
    \begin{align*}
        f_j(t) = 0 \quad &\Longleftrightarrow\qquad  t< \tau_j \quad \Longleftrightarrow\qquad m - t > m - \tau_j.
    \end{align*}
    Recall that the $(2j-1,2j)$-coordinates of $\phi(j,t)$ is $(m-t, t)$. The second condition is equivalent to $\phi(j,t) \in B_0$, while the third is equivalent to $\phi(j,t) \notin B_1$. Thus, $f$ is a 2-box partition.
        
    \paragraph{Generalization to $k \geq 2$.} We now make $k/2$ translated copies of the 2-box construction with $m \leq \frac{4n}{kd}$, so that there are a total of $k$-classes. Define the point set as
    \[X = \bigcup_{i \in [k/2]} \bigcup_{j \in [d/2]} \phi_j([m]) + (mi)\mathbf{1},\]
    where in the $i$th copy, we translate each point by the vector $(mi,\ldots,mi)$. As before, each point in $X$ corresponds to a unique set of indices $(i,j, t) \in [k/2] \times [d/2] \times [m]$. Letting $\phi : (i,j,t) \mapsto x$ denote this bijection, we define $f$ similarly, where $f(x) = 2i + f_{i,j}(t)$ when $x = \phi(i,j,t)$. In this way, we obtain an embedding of $C$ independent threshold problems and the stated lower bound.
\end{proof}

\section{Disjoint Box Partitions} \label{sec:disjoint-box-partitions}

In this section we prove our results for disjoint $k$-box partitions stated in \Cref{thm:disjoint}. We start by proving our lower bound in \Cref{sec:LB-disjoint} for the large-$k$ regime, then obtain a separation between disjoint and general boxes in \Cref{sec:disjoint-alg} with an improved algorithm for the small-$k$ regime.

\subsection[Matching Lower Bound in the Large-k Regime]{Matching Lower Bound in the Large-$k$ Regime} \label{sec:LB-disjoint}

We prove the following lower bound for learning disjoint $k$-box partitions. In particular, we obtain a $\Omega(kd \log(n/kd))$ query lower bound for fully adaptive, randomized algorithms when $k \geq 2^d$, showing that when $k$ is sufficiently large, disjointness does not make the learning problem easier. 

\begin{theorem}[Lower bound for disjoint boxes]
    \label{thm:disjoint-LB}
    Let $k \geq 2$ be a power of two and $d \geq 1$. Define $L = \min(d,\log_2 k)$ and let $2 \leq m \leq \frac{2n}{kL}$ be a natural number. Suppose a randomized algorithm learns any disjoint $k$-box partition over $n$ points in $\mathbb{R}^d$ with success probability $2/3$. Then, there exists a problem instance such that the expected number of queries made is at least $\Omega(kL \log m)$. Moreover, if the algorithm uses $1 \leq r \leq \log_2 m$ rounds of adaptivity, then there exists a problem instance such that the expected number of queries made is at least $\Omega(rkL m^{1/r})$.
\end{theorem}

\begin{proof} We first prove the theorem when $k = 2^d$, and thus $L = d$. The more general case will follow easily from this setting. For the sake of simplicity, assume that $kd$ divides $2n$. We will construct a set $X \subseteq \mathbb{R}^d$ of $n$ points and a family of $k = 2^d$ boxes that induce $kd/2$ independent threshold learning problems, each on a set of $m = 2n/dk$ points. The lower bound will then follow by applying \Cref{thm:LB-threshold} across each independent threshold problem.

Our construction is built on the $d$-dimensional hypercube, $\{0,1\}^d$. Each vertex $u \in \{0,1\}^d$ corresponds to a label induced by a box $B_u$ such that $u \in B_u \subseteq [0,1]^d$. These boxes are disjoint. 

\paragraph{Notation.} Given $u \in \{0,1\}^d$ and $p \in [d]$, we let $u_{-p} \in \{0,1\}^{d-1}$ denote the \emph{context vector} obtained by removing the $p$-th bit of $u$. For a context vector $z \in \{0,1\}^{d-1}$ and coordinate $p \in [d]$, we use the pair $(p,z)$ to identify the hypercube edge with context $z$ in direction $p$.

\begin{figure}[t]
    \centering
    \begin{tikzpicture}[
    line cap=round,
    line join=round,
    scale=0.5,
    edgept/.style={circle, fill=black, inner sep=1pt}
]
\def\L{8}
\def\xa{4.7}
\def\ya{4.5}
\def\xb{3.3}
\def\yb{3.3}

\draw[thick] (0,0) rectangle (\L,\L);

\filldraw[
    fill=blue,
    fill opacity=0.25,
    draw=gray!60!gray,
    draw opacity=1] (0,\L) rectangle (\xa,\yb);
\filldraw[
    fill=orange,
    fill opacity=0.25,
    draw=gray!70!gray,
    draw opacity=1
] (\xb,0) rectangle (\L,\ya);

\draw[thick,gray] (0,0) rectangle (\L,\L);

\foreach \t in {1,2,...,7} {
    \node[edgept] at (\t,0) {};
    \node[edgept] at (\t,\L) {};
    \node[edgept] at (0,\t) {};
    \node[edgept] at (\L,\t) {};
}
\end{tikzpicture}\hspace{3em}
    \begin{tikzpicture}[
    line cap=round,
    line join=round,
    scale=0.5,
    edgept/.style={circle, fill=black, inner sep=1pt}
]
\def\L{8}
\def\xa{4.7}
\def\ya{4.5}
\def\xb{3.3}
\def\yb{3.3}

\draw[thick] (0,0) rectangle (\L,\L);

\filldraw[
    fill=blue,
    fill opacity=0.25,
    draw=gray!60!gray,
    draw opacity=1] (0,\L) rectangle (0.245*\L,0.215*\L);

\filldraw[
    fill=green,
    fill opacity=0.25,
    draw=gray!60!gray,
    draw opacity=1] (0,0) rectangle (0.845*\L,0.215*\L);

\filldraw[
    fill=orange,
    fill opacity=0.25,
    draw=gray!60!gray,
    draw opacity=1] (0.845*\L,0) rectangle (\L,0.785*\L);

\filldraw[
    fill=violet,
    fill opacity=0.25,
    draw=gray!60!gray,
    draw opacity=1] (0.245*\L,\L) rectangle (\L,0.785*\L);

\draw[thick,gray] (0,0) rectangle (\L,\L);

% Top Left
\node[edgept] at (0.17*\L,\L) {};
\node[edgept] at (0.2*\L,\L) {};
\node[edgept] at (0.23*\L,\L) {};
\node[edgept] at (0.26*\L,\L) {};

% Bottom Left
\node[edgept] at (0,0.17*\L) {};
\node[edgept] at (0,0.2*\L) {};
\node[edgept] at (0,0.23*\L) {};
\node[edgept] at (0,0.26*\L) {};

% Bottom Right
\node[edgept] at (0.77*\L,0) {};
\node[edgept] at (0.8*\L,0) {};
\node[edgept] at (0.83*\L,0) {};
\node[edgept] at (0.86*\L,0) {};

% Top Right
\node[edgept] at (\L,0.77*\L) {};
\node[edgept] at (\L,0.8*\L) {};
\node[edgept] at (\L,0.83*\L) {};
\node[edgept] at (\L,0.86*\L) {};

\end{tikzpicture}
    \caption{A visualization for the construction in \Cref{thm:disjoint-LB}. The left and right subfigures give two point configurations on the edges of the two-dimensional Boolean hypercube. The left configuration fails to embed a set of independent threshold problems, since thresholds induced by overlapping boxes, like the one shown, are immediately ruled out. The point configuration on the right embeds an independent set of threshold problems per edge using a coherent family of total orderings over context vectors. In particular, the horizontal edges are ordered top-to-bottom and the vertical edges are ordered left-to-right.}
    \label{fig:LB-disjoint}
\end{figure}
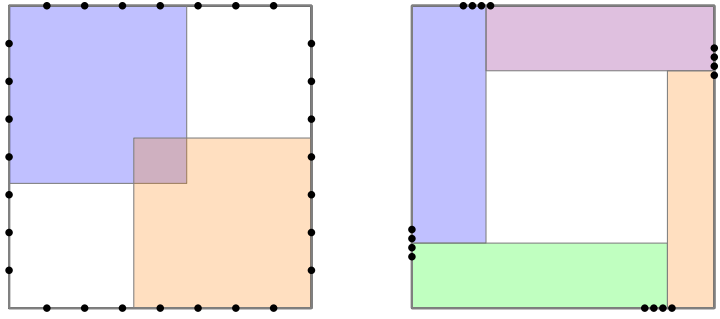

\paragraph{The construction.} Our objective is to induce an independent threshold learning problem along each edge $(p,z)$. For each dimension $p \in [d]$ and context $z \in \{0,1\}^{d-1}$, we define a disjoint family of intervals as follows:
\begin{align} \label{eq:interval-partition}
    \mathcal{I}(p) = \Big\{I(p,z) \subset [0,1] \colon z \in \{0,1\}^{d-1}\Big\} \text{ such that } \bigsqcup_{z \in \{0,1\}^{d-1}} I(p,z) = [0,1] \text{.}
\end{align}
These intervals are of uniform length, $2^{-(d-1)}$. Note that what is currently left undefined is the \emph{ordering} of the intervals according to their context vectors, $z$. This turns out to be a critical aspect of the construction, which we will discuss and define later.

On each edge $(p,z) \in [d] \times \{0,1\}^{d-1}$, we induce a threshold learning problem on the segment
\[
L(p,z) = \Big\{ x \in [0,1]^d \colon x_{-p} = z \text{ and } x_p \in I(p,z)\Big\} \text{.}
\]
To do this, we let $X(p,z) \subset L(p,z)$ be a set of $\frac{n}{d2^{d-1}} = \frac{2n}{kd} = m$ uniformly spaced interior points along this segment. Concretely, if $I(p,z) = [a,a+h]$, then define 
\[
\delta = \frac{h}{2m+2} ~\text{ and }~ X(p,z) = \Big\{x \in [0,1]^d \colon x_{-p} = z, x_p \in \{a+2i\delta \colon 1 \leq i \leq m\}\Big\} \text{.}
\]
We define the full point-set as 
\[
X = \bigcup_{p \in [d], z \in \{0,1\}^{d-1}} X(p,z) .
\]
A threshold $\tau(p,z) \in I(p,z)$ is then chosen somewhere in this interval. Concretely,
\[
\tau(p,z) \in \Big\{a + (2i+1)\delta \colon 0 \leq i \leq m\Big\}
\] is one of $m+1$ possible values. Then, given a threshold profile
\[
T = \Big\{\tau(p,z) \colon p \in [d], z \in \{0,1\}^{d-1}\Big\}
\]
the box for label $u \in \{0,1\}^d$ is defined as
\[
B_u = \left(\prod_{p \in [d] \colon u_p = 0} [0,\tau(p,u_{-p}) - \delta/2]\right) \times \left(\prod_{p \in [d] \colon u_p = 1} [\tau(p,u_{-p}) + \delta/2, 1]\right).
\]
The additive $\pm \delta/2$ term is simply to ensure that the following closed intervals are disjoint: 
\[[0,\tau(p,u_{-p}) - \delta/2]\quad\textrm{and} \quad [\tau(p,u_{-p}) + \delta/2, 1].\] 
Note that the union of these boxes indeed cover $X$, regardless of the threshold profile $T$. The crux of the construction is to show that all $(m+1)^{kd/2} = (\frac{2n}{kd} + 1)^{kd/2}$ threshold profiles are realizable while maintaining box disjointness, i.e. $B_u \cap B_v = \emptyset$ for all $u \neq v \in \{0,1\}^d$. 

\paragraph{Completing the proof assuming realizability of all threshold profiles.} We first note that if the above is possible, then this completes the proof. Learning each threshold on $X(p,z)$ amounts to learning a monotone function on a domain of size $m = \frac{2n}{kd}$. The needed lower bound for learning each threshold is therefore given by \Cref{thm:LB-threshold} due to \cite{thresholds-LB-ref}. 

\paragraph{Achieving box disjointness for all threshold profiles.} Note that $B_u \cap B_v = \emptyset$ exactly when in some dimension, the corresponding intervals are disjoint. When $u,v$ are adjacent, this holds as
\[
[0,\tau(p,u_{-p}) - \delta/2] \cap [\tau(p,u_{-p}) + \delta/2, 1] = \emptyset \text{.}
\]
However, it is not clear a priori how to make this hold for all $u,v$ where $d(u,v) \geq 2$. Suppose $u_p = 0, v_p = 1$ and $u_q \neq v_q$ for some $q \neq p \in [d]$. A sufficient property to guarantee disjointness is $\tau(p,u_{-p}) < \tau(p,v_{-p})$. More generally, \emph{for every} $u,v$ where $d(u,v) \geq 2$, we need the following to hold: there exists a coordinate $p$ such that $u_p = 0, v_p = 1$ and $\tau(p,u_{-p}) < \tau(p,v_{-p})$. Recall that the threshold $\tau(p,u_{-p}) \in I(p,u_{-p})$ is chosen from the interval partition corresponding to the edge $(p,u_{-p})$ (recall \cref{eq:interval-partition}). The ordering on these intervals has not yet been defined. We now define their ordering in such a way that the desired property is satisfied for all threshold profiles. For each $p \in [d]$, we will define a total order $\prec_p$ over $z \in \{0,1\}^{d-1}$ so that $(\prec_1, \prec_2, \ldots, \prec_d)$ together have the following property.

\begin{definition} [Coherent family of orderings] \label[definition]{def:coherence} Call total orders $(\prec_1, \prec_2, \ldots, \prec_d)$ over context vectors $\{0,1\}^{d-1}$ \emph{coherent} if the following holds. Let $x \neq y \in \{0,1\}^d$ be any two non-adjacent vertices, i.e. $d(x,y) \geq 2$. Then, there is a coordinate $p$ on which they differ such that the context of the $p$-bit-$0$ vertex precedes the context of the $p$-bit-$1$ vertex in $\prec_p$. I.e. if $x_p = 0, y_p = 1$, we have $x_{-p} \prec_p y_{-p}$ and if $x_p = 1, y_p = 0$, we have $y_{-p} \prec_p x_{-p}$. \end{definition}

The existence of a coherent family of total orders $(\prec_1, \prec_2, \ldots, \prec_d)$ (according to \Cref{def:coherence}) completes the proof by the above argument. We now establish the existence of such a family. (See \Cref{fig:LB-disjoint} for an illustration of the final construction in $2$-dimensions.) 

\paragraph{Coherent orderings of context vectors.}

The heart of the construction is the following coherent family of orderings yielding the necessary structure. 

\begin{definition} \label[definition]{def:context-orderings} For each $p\in[d]$, let $\prec_p$ be a total order on the contexts $\{0,1\}^{[d]\setminus\{p\}}$ defined as follows. For each $u,v \in \{0,1\}^{[d]\setminus\{p\}}$, look at the first bit $q$ where they differ. Suppose $u_q = 0$, $v_q = 1$. If $q < p$, then $v \prec_p u$. If $q > p$, then $u \prec_p v$. In other words, $\prec_p$ is a lexicographic order where all bits before $p$ use the reversed bit order and all bits after $p$ use the standard bit order. \end{definition}

\begin{claim} [Coherence property] \label{claim:coherence}
The family of orderings $(\prec_1,\prec_2,\ldots,\prec_d)$ defined in \Cref{def:context-orderings} satisfies the coherence property of \Cref{def:coherence}. \end{claim}

\begin{proof} Let $a < b$ be the first two bits where $x$ and $y$ differ. First, suppose that $x_{ab} = 00$, $y_{ab} = 11$ (the opposite case is symmetric). Observe that $b$ is the first bit on which the context vectors $x_{-a}$ and $y_{-a}$ disagree. Since $a < b$, we then have $x_{-a} \prec_a y_{-a}$ by definition of the ordering $\prec_a$.

Now, suppose $x_{ab} = 01, y_{ab} = 10$ (again, the opposite case is symmetric). Observe that $a$ is the first bit on which the context vectors $x_{-b}$ and $y_{-b}$ disagree. Since $a < b$, we then have $y_{-b} \prec_b x_{-b}$ by definition of the ordering $\prec_b$. \end{proof}

\paragraph{Generalizing beyond the $k = 2^d$ case.} We have completed the proof for the case when $k = 2^d$. We now extend beyond this setting. For simplicity, we will assume $k = 2^{\ell}$ is some power of $2$.
\begin{itemize}
    \item $\ell < d$. In this case, we can simply apply the previous construction in $\ell$-dimensions and pad the instance with $d-\ell$ dummy coordinates. We have $\ell = \log_2 k = \min(d,\log_2 k)$, establishing the lower bound claimed in \Cref{thm:disjoint-LB}.
    \item $\ell > d$. In this case, we create $2^{\ell - d}$ copies of the previous construction in $d$ dimensions, translated along the first coordinate. The $n$ points are divided evenly among these copies. It can be easily verified that this again results in the lower bound claimed in \Cref{thm:disjoint-LB}.
\end{itemize}

\noindent This completes the proof of \Cref{thm:disjoint-LB}. \end{proof}

\subsection[Beating the General Box Lower Bound in the Small-k Regime]{Beating the General Box Lower Bound in the Small-$k$ Regime} \label{sec:disjoint-alg}

In \Cref{sec:general-box-partitions}, we saw that learning a labeling induced by $k$ boxes can require at least $\Omega_k(d \log (n/d))$ queries when $k$ is constant order. These boxes were allowed to overlap, although no data points may lie in the overlapping region. It turns out that if the boxes are guaranteed to be induced by \emph{disjoint} boxes, then in the small-$k$ regime, it is possible to beat the lower bound. The next result, \Cref{thm:disjoint-small-UB}, shows that when $k$ is constant order, there is an upper bound of $O_k(d \log(n)/\log(d))$ queries for learning the restricted class of disjoint boxes, which shaves off a factor of $\log(d)$.

The algorithm that achieves this faster rate is a variant of a classic learning algorithm, which we refer to as Halving++, described in \cite{DBLP:conf/colt/Hegedus95}. Compared to CAL, Halving++ selects query points more carefully by searching for more informative points to query. It turns out that the more methodical selection is needed; we show in \Cref{sec:separation} that there are disjoint-box problems where CAL requires $\Omega(kd \log (n/kd))$ queries to learn, strictly separating CAL and Halving++.

\begin{theorem} [Disjoint Boxes] \label{thm:disjoint-small-UB}
    Let $d, k \geq 2$. There is a deterministic algorithm that learns the labeling of a disjoint $k$-box partition over a worst-case set of $n$ points in $\mathbb{R}^d$ using
    \[O\left(\min\left\{n, \frac{k^3 d \log n}{\log (kd)}\right\}\right) \textrm{ queries.}\]
\end{theorem}

\begin{proof}
    The proof of this uses a classic learning theory result bounding the query complexity through the \emph{extended teaching dimension} of a binary concept class \cite{moshkov1982conditional,DBLP:conf/colt/Hegedus95}.

    \begin{definition}[Extended teaching dimension]
        Let $Z$ be a finite set and let $\mathcal{C} \subset \{0,1\}^Z$. For any $h \in \{0,1\}^Z$, a subset $S \subset Z$ is a \emph{specifying set} for $h$ with respect to the concept class $\mathcal{C}$ if at most one element $f \in \mathcal{C}$ satisfies $f(z) = h(z)$ for all $z \in S$. The \emph{extended teaching dimension} of $\mathcal{C}$ is
        \[\mathrm{ETD}(\mathcal{C}) = \max_{h \in \{0,1\}^Z}\min \big\{|S| : S \textrm{ is a specifying set for $h$ with respect to $\mathcal{C}$}\big\}.\]
    \end{definition}

    \begin{theorem}[Query complexity bound, \cite{moshkov1982conditional,DBLP:conf/colt/Hegedus95}] \label{thm:moskhov-hegedus}
        Let $\mathcal{C} \subset \{0,1\}^Z$ be a concept class. If $\mathrm{ETD}(\mathcal{C}) \geq 2$ and $\max\{3, \mathrm{ETD}(\mathcal{C})\} \leq E$, there is a deterministic algorithm that identifies any concept in $\mathcal{C}$ using at most
        \[\hspace{2em}\frac{2 E \log_2 |\mathcal{C}|}{\log_2 E}  \textrm{ adaptive coordinate queries.}\]
    \end{theorem}

    First, we convert the problem of learning a $k$-box partition into one of learning a binary concept class. Let $Z = X \times [k]$. Each labeling $f : X \to [k]$ is associated to an element of $\tilde{f} \in \{0,1\}^Z$
    \[\tilde{f}(x, y) = 1 \quad \Longleftrightarrow \quad f(x) = y.\]
    Let $\mathcal{C}_{X}^{\textrm{$k$-DISJ}} \subset \{0,1\}^Z$ denote the concept class 
    \begin{equation}\label{eqn:kdisj}
    \mathcal{C}_{X}^{\textrm{$k$-DISJ}} = \{\tilde{f} : f \textrm{ is a disjoint $k$-box partition on $X$}\}.
    \end{equation}
    To recover the ground truth labeling $f: X \to [k]$, it is enough to recover its associated concept $\tilde{f}$. Every coordinate query to $z = (x, y)$ can be simulated by a single label query to $x$, so it follows that by running the algorithm from \Cref{thm:moskhov-hegedus} as a black-box, we can learn any $f$ by making at most
    \[\hspace{2em}\min\left\{n, \frac{2 E \log_2 |\mathcal{C}_{X}^{\textrm{$k$-DISJ}}|}{\log_2 E} \right\} \textrm{ adaptive label queries},\]
    where $E \geq 3$ is any upper bound on the extended teaching dimension. The result follows immediately from \Cref{lem:etd-bound,lem:disj-count} below, which bound $\mathrm{ETD}(\mathcal{C}_{X}^{\textrm{$k$-DISJ}})$ and $|\mathcal{C}_{X}^{\textrm{$k$-DISJ}}|$.
\end{proof}

    \begin{lemma} \label[lemma]{lem:etd-bound}
        Let $\mathcal{C}_{X}^{\textrm{$k$-$\mathrm{DISJ}$}}$ be the class of disjoint $k$-box partitions on $X \subset \mathbb{R}^d$ defined in \eqref{eqn:kdisj}. Then:
        \[\mathrm{ETD}(\mathcal{C}_{X}^{\textrm{$k$-$\mathrm{DISJ}$}}) \leq 2kd.\]
    \end{lemma}
    \begin{proof}
        Fix any $h \in \{0,1\}^Z$. While $h$ is not necessary associated with a disjoint $k$-box labeling, but we can still interpret it as assigning each point $x \in X$ to a possibly empty set of labels, where the assignment is $x \mapsto \{y \in [k]: h(x,y) = 1\}$. There are three cases:
        \begin{itemize}
            \item If there is an $x \in X$ assigned zero labels or strictly more than one label, then the following set $S = \{(x, y): y \in [k]\}$ is a specifying set for $h$ with respect to $\mathcal{C}$. No $\tilde{f} \in \mathcal{C}_{X}^{\textrm{$k$-$\mathrm{DISJ}$}}$ such that $\tilde{f}(x,y) = h(x,y)$ for all $x \in S$ and $y \in [k]$.
            \item Suppose that each $x \in X$ is assigned precisely one label. For each $y \in [k]$, let $B_y$ be the bounding box of points assigned the label $y$ by $h$. In each dimension of this bounding box, we can let $x_{y, i, L}$ and $x_{y, i, R}$ be the left- and rightmost points out of all points labeled $y$. Define $S$ to be the union of these bounding points, so that $S$ is a set of size at most $2kd$. There is precisely one (not necessarily disjoint) box partition that induces the same labeling on $S$.
        \end{itemize}
        By definition, the extending teaching dimension of $\mathcal{C}_{X}^{\textrm{$k$-$\mathrm{DISJ}$}}$ is at most $2kd$.
    \end{proof}

    \begin{lemma}\label[lemma]{lem:disj-count}
        There are at most $(2kdn)^{\binom{k}{2}}$ disjoint $k$-box partitions on any $n$-point set $X \subset \mathbb{R}^d$.
    \end{lemma}

    \begin{proof}
        Define $\Theta = X \times [d] \times \{-1, +1\}$. We will construct a family of test functions $t_\theta : X \to \{0,1\}$ parametrized by $\theta \in \Theta$ such that the following holds: for any disjoint $k$-box labeling $f : X \to [k]$ and pair of distinct labels $y_0, y_1 \in [k]$, there is a test function $t_\theta$ such that
        \[f(x) = y_0 \implies t_\theta(x) = 0\qquad\textrm{and}\qquad f(x) = y_1 \implies t_\theta(x) = 1.\]
        Thus, $f$ is completely determined by a set of $\binom{k}{2}$ parameters $\{\theta(f,\ell_0, \ell_1) : 1 \leq \ell_0 < \ell_1 \leq k\}$, and
        \[|\mathcal{C}_{X}^{\textrm{$k$-$\mathrm{DISJ}$}}| \leq |\Theta|^{\binom{k}{2}}.\]
        The claim holds because $|\Theta| = 2dn$. We just need to construct the family of test functions.
        
        For any labeling $f: X \to [k]$ and pair of distinct labels $y_0, y_1 \in [k]$, let $B_0$ and $B_1$ respectively be the bounding boxes of points in class $y_0$ and $y_1$. These two boxes are disjoint if and only if there exists a separating dimension $i \in [d]$ such that the projection of these boxes onto the $i$th dimension is two disjoint intervals. Let $x \in X \cap B_1$ minimize the distance to $B_0$ after projecting to the $i$th dimension
        \[\min_{x_0 \in B_0} \, |x_0(i) - x(i)| \leq \min_{x_j \in X \cap B_j}|x_0(i) - x_1(i)|.\]
        Let $s = \mathrm{sign}\big(x(i) - x_0(i)\big)$ for any $x_0 \in B_0$. This is well-defined because $x(i)$ is either strictly greater or strictly less than the $i$th coordinate of every $x_0 \in B_0$. A simple test to distinguish points in class $\ell_0$ and $\ell_1$ is the following test parametrized by $\theta = (i, x, s)$
        \[t_\theta(x') = \mathbf{1}\big\{s \cdot x'(i) \geq s \cdot x(i)\big\}.\]
        By construction, $t_\theta(x_0) = 0$ for all $x_0 \in B_0$ and $t_\theta(x_1) = 1$ for all $x_1 \in B_1$. 
    \end{proof}

\begin{remark}[Applying the ETD argument to general boxes]
    The same argument can be applied to the earlier, general setting of learning $k$ potentially-overlapping boxes. The proof of \Cref{lem:etd-bound}, which gave a bound of $2kd$ on the extended teaching dimension, also applies to general boxes; nowhere was disjointness used. The bound on the number of $k$-box partitions can be upper bounded by $n^{2kd}$, each box can be identified with two endpoints in each dimension. Then, \Cref{thm:moskhov-hegedus} yields
    \[\hspace{1em}\frac{8 k^2 d^2 \log n }{\log (2kd)} \textrm{ queries.}\]
    This upper bound is worse than \Cref{thm:overlap}, although it is likely that the analysis is loose. But, for disjoint boxes, the savings achieved in \Cref{thm:disjoint-small-UB} intuitively relies the fact that we can distinguish those boxes by a set of $\Theta(k^2)$ separating hyperplanes, rather than their $\Theta(kd)$ endpoints.
\end{remark}

\section{Decision Trees}\label{sec:decision-tree}

In this section we prove our results for decision trees stated in \Cref{thm:dt}. 

\begin{definition} [Decision trees] \label[definition]{def:decision} Let $Y$ be a label set of size $|Y| = k \geq 1$. A function $f \colon \mathbb{R}^d \to Y$ is a $k$-leaf \emph{decision tree} if either (a) $k = 1$, or (b) there exists $i \in [d]$, $\tau \in \mathbb{R}$, and a strict subset $\emptyset \neq S \subsetneq Y$ such that the following hold:
\begin{itemize}
    \item $f$ restricted on $\{x \in \mathbb{R}^d \colon x_i \leq \tau\}$ is a $|S|$-leaf decision tree with label set $S$ and 
    \item $f$ restricted on $\{x \in \mathbb{R}^d \colon x_i > \tau\}$ is a $(k-|S|)$-leaf decision tree with label set $Y \setminus S$.
\end{itemize}
A $k$-leaf \emph{decision list} is defined analogously, with the added restriction that $|S| = 1$ is a singleton. \end{definition}

A decision stump is a $2$-leaf decision tree. Our results in this section use the following learner for decision stumps as a black-box. 

\begin{theorem} [Corollary 2 item (1) of \cite{DBLP:conf/colt/BlackLMSS26}] \label{thm:stump-learn} Let $X \subset \mathbb{R}^d$ be an arbitrary set of $n$ points. There exists a deterministic algorithm, that exactly learns an arbitrary decision stump $f \colon X \to \{0,1\}$ using $O(d + \log n)$ queries. If the input $f$ is not realizable by a decision stump, the algorithm still halts within the claimed query budget and returns some labeling $g \colon X \to \{0,1\}$. \end{theorem}

We use $\mathsf{stumpLearn}(f,X)$ to denote the algorithm of \Cref{thm:stump-learn}. To see that the claim holds regarding inputs not realizable by decision stumps, one can simply truncate the learner at its promised worst-case query depth. If the learner is not done by this point, complete the labeling arbitrarily and return the result. Of course, correctness is only guaranteed in the realizable case. 

\subsection{Decision Tree Learner} 

We prove the following theorem, obtaining a $2^{O(k)}(d+\log n)$ upper bound.

\begin{theorem} \label{thm:DT-learner} Let $Y$ be an arbitrary label-set of size $|Y| = k \geq 1$. There is a deterministic algorithm, which for any set $X \subset \mathbb{R}^d$ of $n$ points, learns an unknown labeling $f \colon X \to Y$ induced by a $k$-leaf decision tree using 
\[
2^{2k + O(\log^2 k)} (d+ \log n) = 2^{O(k)} (d+ \log n)
\]
queries. If $f$ is induced by a $k$-leaf \emph{decision list}, then the query complexity is $O(k^2(d+ \log n))$. \end{theorem}

Before proving \Cref{thm:DT-learner}, we remark that it is fairly direct to obtain a $k^{O(k)}(d+ \log n)$ learner by enumerating all decision trees over the label set and applying $\mathsf{stumpLearn}$ at every internal node. 

\begin{remark} [A simple tree-enumeration algorithm] \label[remark]{rem:DT-simple} We give a sketch of a $k^{O(k)}(d+\log n)$ query algorithm. Let $T$ be a binary tree with $k$ leaves where every non-leaf has exactly two children. Note that there are exactly $C_{k-1}$ (the $(k-1)$-th Catalan number) choices for $T$. Then, let $\phi$ be a bijective map from the label set $Y$ to the leaves of $T$ (there are $k!$ such maps). For a node $u$ in $T$, let $Y_u \subseteq Y$ denote the set of labels associated by $\phi$ to some leaf-descendant of $u$. Now, for each of the $k-1$ non-leaf nodes $u$, run $\mathsf{stumpLearn}$ (\Cref{thm:stump-learn}) with Boolean oracle $g_u(x) = \mathbf{1}(f(x) \in Y_{\mathsf{right}_T(u)})$. Combining the outputs of these calls to the stump learner yields a candidate labeling $f_{T,\phi} \colon X \to Y$. By \Cref{def:decision}, there exists at least one choice of $(T,\phi)$ such that $f_{T,\phi} = f$. It is then straightforward to determine which of these $C_{k-1}k!$ candidate labelings is the correct one using at most one query per candidate. The total query complexity is 
\[
O(C_{k-1}k! k(d+ \log n)) = k^{O(k)}(d+\log n).
\]
\end{remark}

Our improvement over the brute-force enumeration described in \Cref{rem:DT-simple} comes from the following two observations:
\begin{itemize}
    \item There is a simple procedure (\Cref{clm:mono-cert}) to certify correctness at the leaves, i.e. when the label-set is a singleton. This allows us to make permanent progress (by labeling one class) without needing to guess the entire correct binary tree over the labels all at once.
    \item Given label-set $Y$, there exists $\emptyset \subsetneq A^{\star} \subsetneq Y$ where the root-level partition on the labels is $(A^{\star}, Y \setminus A^{\star})$ such that $|A^{\star}| \leq \floor{|Y|/2}$. As a result, the recurrence governing the query complexity has the form $Q(k) \lesssim 2^kQ(k/2)$. 
\end{itemize}

\paragraph{Decision tree learner.} We now describe the algorithm, which uses the following subroutine in addition to the stump learner of \Cref{thm:stump-learn}.

\begin{claim} [Monochromatic Certification] \label{clm:mono-cert} Let $f \colon X \to Y$ be a labeling induced by a box partition, $Z \subseteq X$, and $y \in Y$. There is a deterministic procedure $\mathsf{monoCert}(f,Z,y)$ using at most $2d$ queries that decides whether or not $f|_{Z} \equiv y$. \end{claim}

\begin{proof} The procedure simply queries all points in the set
\[
E_Z = \left\{\argmin_{z \in Z} z_i, \argmax_{z \in Z} z_i \colon i \in [d] \right\}
\]
of extremal points in $Z$ with respect to each coordinate. Clearly $|E_Z| \leq 2d$. We then accept if and only if $f|_{E_Z} \equiv y$. Clearly, if $f|_{E_z} \not\equiv y$, then $f|_{Z} \not\equiv y$. Conversely, since $Z \subseteq \mathsf{bbox}(E_Z)$, if $f|_{E_z} \equiv y$, then $f|_{Z} \equiv y$. \end{proof}

\noindent Given $f \colon X \to Y$, our algorithm is defined recursively as follows. 

\begin{mdframed}
\noindent $\mathsf{DTLearn}(f,X,Y)$:
\begin{enumerate}
    \item \textbf{Base cases.} If $X = \emptyset$, return the empty labeling. If $Y = \{y\}$, run $\mathsf{monoCert}(f,X,y)$ (\Cref{clm:mono-cert}). If it accepts, return the constant labeling $f|_{X} \equiv y$, otherwise return $\mathsf{FAIL}$. 
    \item If $|Y| = k \geq 2$, then let $R \gets X$ and repeat the following procedure defined in item (3) $k$ times, or until $R = \emptyset$. (See line (3e).) 
    
    \item For every $A \subset Y$ where $1 \leq |A| \leq \floor{k/2}$, do the following:
    \begin{enumerate}
        \item Define $g_A \colon R \to \{0,1\}$ where $g_A(x) = \mathbf{1}(f(x) \in A)$.
        \item Run $\mathsf{stumpLearn}(g_A,R)$ and let $\widetilde{g}_A$ denote the returned Boolean labeling. 
        \item Let $X_A = \{x \in R \colon \widetilde{g}_A(x) = 1\}$.
        \item Run $\mathsf{DTLearn}(f|_{X_A}, X_A,A)$. If it returns $\mathsf{FAIL}$, then continue. If it returns a labeling $h_A \colon X_A \to A$, then assign these labels to $X_A$ and set $R \gets R \setminus X_A$. 
        \item If $R = \emptyset$, return the accumulated labeling on the original input set.
    \end{enumerate}
    \item If after $k$ repetitions of item (3), we have $R \neq \emptyset$, then return $\mathsf{FAIL}$.
\end{enumerate}
\end{mdframed}

Before proving correctness, we give some intuition for the above learner. Since $f$ is a decision tree, for some choice of $A$ in line (3), the Boolean labeling $g_A$ in line (3a) is a decision stump. By correctness of the decision stump learner, this implies that $f|_{X_A}$ is a $|A|$-leaf decision tree and thus by induction the points in $X_A$ are correctly labeled in line (3d). This removes at least one of the $k$ classes from $R$, which is the reason that at most $k$ repetitions of (3) are needed. Importantly, even when $g_A$ is not a decision stump, any labeling returned by the recursive call is correct, preventing us from ever mislabeling points.

\subsubsection{Correctness Proof}

\begin{lemma} [Soundness] \label{lem:soundness} Suppose $f \colon X \to Y$ satisfies $f^{-1}(y) = X \cap B_{y}$ where $\{B_{y} \colon y \in Y\}$ are disjoint axis-parallel boxes. Let $Y'$ be a candidate label set for $f$. If $\mathsf{DTLearn}(f,X,Y')$ returns a labeling $h$, then $f = h$. \end{lemma}

\begin{proof} Every label assignment returned by $\mathsf{DTLearn}$ ultimately comes from a successful monochromatic certification at a singleton base case. \Cref{clm:mono-cert} guarantees that every such assignment is correct, regardless of whether the candidate label set $Y'$ contains all true labels. A complete labeling is returned only after every point has received a label. \end{proof}

\begin{lemma} [Correctness] \label[lemma]{lem:correctness} Suppose $f \colon X \to Y$ is induced by a $\ell$-leaf decision tree where $\ell \leq |Y|$. Then $\mathsf{DTLearn}(f,X,Y)$ always returns the correct labeling $f$. \end{lemma}

\begin{proof} The proof is by induction on $k = |Y|$. The case $X = \emptyset$ is clear and the case $k = 1$ follows from \Cref{clm:mono-cert}. Now, let $k \geq 2$. Consider the set $R$ defined in line (2) and consider some iteration of step (3) in which $R \neq \emptyset$. We argue that one iteration of step (3) correctly labels all points in at least one of the $k$ labels. (Therefore, $k$ repetitions are enough, as claimed in line (2).) Note that by \Cref{lem:soundness}, a recursive call will never incorrectly label points. 

First, suppose $f|_R \equiv y$ is constant, for some $y \in Y$. Then, in step (3), the algorithm will try $A = \{y\}$, in which $g_{\{y\}} \equiv 1$ (in line (3a)). This will result $\mathsf{stumpLearn}(g_{\{y\}},R)$ returning the all-$1$ labeling of $R$ and sets $X_A = R$ in line (3c). Line (3d) then invokes the base case, accepts, and labels all of $R$. 

Now, suppose $R$ contains at least $2$ labels from $Y$. Note that $f|_R \colon R \to Y$ is itself induced by a $k'$-leaf decision tree where $k' \leq k$ (this holds for any subset of $X$). Therefore, there exists some $A^{\star} \subset Y$ where $1 \leq |A^{\star}| \leq \floor{k/2}$ such that either $f(x) \in A^{\star}$ iff $x_i \leq \tau$ (for some $i,\tau$) or $f(x) \in A^{\star}$ iff $x_i > \tau$. In some iteration of line (3), the algorithm tries this $A^{\star}$. The guarantee of $\mathsf{stumpLearn}(g_{A^{\star}},R)$ (\Cref{thm:stump-learn}) implies that $\widetilde{g}_{A^{\star}}  = g_{A^{\star}}$ and thus $X_{A^{\star}} = \{x \in R \colon f(x) \in A^{\star}\}$. By induction, the recursive call to $\mathsf{DTLearn}(f_{X|_{A^{\star}}},X_{A^{\star}},A^{\star})$ successfully returns a labeling on $X_{A^{\star}}$, which by \Cref{lem:soundness} is guaranteed to be correct. Therefore, at the end of step (3d) all points with labels in $A^{\star}$ are removed from $R$.

We have shown that every repetition of line (3) successfully learns at least one of the $k$ labels. Combining with \Cref{lem:soundness}, this shows that after at most $k$ repetitions, the algorithm returns the correct labeling, $f$. \end{proof}

\subsubsection{Query Complexity Proof}

We derive the query complexity of $\mathsf{DTLearn}$ as follows. Fix $d$ and $n$. Let $q_s = O(d + \log n)$ denote the query complexity of the stump learner in $d$-dimensions over $n$ points and let $q_c = 2d$ denote the query complexity of the monochromatic certification. For $k \geq 1$, let $Q(k)$ denote the query complexity of $\mathsf{DTLearn}$ on $n$ points in $d$ dimensions with $k$ labels. Then, by inspection, $Q$ satisfies the following recurrence relation:
\begin{align}
    Q(1) \leq q_c ~\text{ and }~ Q(k) \leq k 2^k (q_s + Q(\floor{k/2})) \text{ when } k \geq 2\text{.}
\end{align}
We now bound the recurrence, which establishes the claimed query complexity. We can expand the recurrence for $k\geq 2$ as follows:
\begin{align}
    Q(k) &\leq k 2^k (q_s + Q(\floor{k/2})) \nonumber \\
    &\leq k 2^k q_s + k2^k\left(\frac{k}{2}\cdot2^{k/2}(q_s + Q(\floor{\floor{k/2}/2}))\right) \nonumber \\
    &\vdots \nonumber \\ 
    &\leq (q_s + q_c) \sum_{t=0}^{\lfloor \log_2 k \rfloor} \left( \prod_{j=0}^t \frac{k}{2^j} 2^{k/2^j} \right) 
\end{align}
Using upper bounds $t \leq \lfloor \log_2 k \rfloor$ and $k/2^j \leq k$ shows that the RHS above is at most
\begin{align}
    (q_s + q_c) \cdot (\log_2 k + 1) \cdot \prod_{j=0}^{ \lfloor \log_2 k \rfloor } k 2^{k/2^j} \leq (q_s + q_c) (\log_2 k + 1)\cdot k^{\log_2 k + 1} \cdot 2^{2k} \leq 2^{2k + O(\log^2 k)} \cdot (q_s + q_c)
\end{align}
and this completes the proof.

\paragraph{Quadratic bound for decision lists.} We now describe how to modify $\mathsf{DTLearn}$ to get an improved query complexity for decision lists (recall \Cref{def:decision}). The simple change is to modify step (3) to try only singleton sets for $A$. The algorithm is correct by the same arguments in \Cref{lem:soundness} and \Cref{lem:correctness}. The algorithm correctly learns the labeling simply after $k$ repetitions of step (3) and only ever recurses on a singleton set of labels. Therefore, by inspection the query complexity is at most $k^2 (q_s + q_c)$.

\subsection{Lower Bound}

We now prove an elementary lower bound establishing optimality of our algorithm for constant $k$.

\begin{theorem} [Decision List Lower Bound] Let $d, k\geq 2$. There exists set $X \subset \mathbb{R}^d$ of $n \geq d(k-1)$ points such that any randomized adaptive algorithm that learns a labeling $f \colon X \to [k]$ induced by a $k$-leaf decision list with success probability at least $2/3$ must use $\Omega(k(d+\log(n/k)))$ queries. \end{theorem}

\begin{proof} It suffices to prove an $\Omega(kd)$ lower bound and a $\Omega(k \log (n/k))$ lower bound separately. 

\paragraph{The $\Omega(kd)$ bound.} We first describe the construction for $n = d(k-1)$ points. Let $U = \{e_1,\ldots,e_d\}$ denote the standard unit basis vectors. For a vector $v$, let $U + v = \{u + v\colon u \in U\}$. For a scalar $c \in \RR$, let $\overline{c} = (c,c,\ldots,c) \in \RR^d$. Our point-set $X$ is defined
\[
X := \bigcup_{y = 1}^{k-1} (U + \overline{y}) \text{.}
\]
Then, we consider the family of decision lists formed as follows. For each $y \in [k-1]$, choose a dimension $i_{y} \in [d]$. Given $\alpha =(i_1,\ldots,i_{k-1}) \in [d]^{k-1}$, we then define $f_{\alpha} \colon \mathbb{R}^d \to [k]$ as 
\[
f_{\alpha}(x) = \min \left(\left\{y \in [k-1] \colon x_{i_y} \leq y + 0.5 \right\} \cup \{k\} \right)\text{.}
\]
To reach any $n \geq d(k-1)$ points, we can pad the point-set with $n - d(k-1)$ arbitrary points $z$ such that $z_i > k$ for all $i \in [d]$. 

Observe that such a labeling is a valid decision list (\Cref{def:decision}). The key property of this construction is the following: for each $y \in [k-1]$, $f(e_{i_{y}} + \overline{y}) = y+1$ and $f(e_{i} + \overline{y}) = y$ for all $i \in [d] \setminus \{i_{y}\}$. Therefore, learning the Boolean labeling in $U + \overline{y}$ requires solving an unstructured search problem on $d$ elements, which requires $\Omega(d)$ queries for any randomized algorithm. Summing over $U + \overline{y}$ for each $y \in [k-1]$ shows that $\Omega(kd)$ queries are needed.

\paragraph{The $\Omega(k \log(n/k))$ bound.} Assume $n$ is a multiple of $k-1$. It suffices to prove a lower bound for learning monotone functions $f \colon [n] \to [k]$. First, when $k=2$, a lower bound of $\Omega(\log n)$ holds by a standard information-theoretic argument, since one needs to find a threshold in one of $n$ possible locations using queries that return a single bit of information.

Now, consider the following family of monotone functions. First, partition $[n]$ into $k-1$ intervals $I_1,\ldots,I_{k-1}$ of length $n/(k-1)$. In interval $I_y$, we allow the function to take values in $\{y,y+1\}$. Learning such a function requires the algorithm to learn the correct \emph{Boolean function} restricted on each $I_y$ and crucially a query to $x \in I_y$ gives \emph{no information} about the function on any other interval. Thus, leveraging the $k=2$ lower bound, we obtain a lower bound of 
\[
(k-1) \cdot \Omega\left(\log \left(\frac{n}{k-1}\right)\right) = \Omega(k \log (n/k))
\]
and this completes the proof. \end{proof}

\section*{AI Disclosure}

ChatGPT 5.6 was used during this work to explore proof ideas. All final proofs and exposition have been written and verified by the authors.

\bibliographystyle{alpha}
\bibliography{biblio}

\appendix

\section{Query Complexity with Round Constraints} \label{sec:low-round-alg}

This section proves \Cref{thm:r-round}. Before we do so, we first prove the following key lemma about labelings that are induced by $k$ boxes. Consider any set of points $X$ with initially unknown labels. This lemma shows that if we label a random subset of points $Z \subseteq X$ of size $O(kdm)$, then in expectation, we will be able to deduce the labels for all but a $O(\frac{1}{m})$-fraction of points in $X$.

\begin{lemma} [Expected coverage of a random set] \label[lemma]{lem:expected-coverage} 
    Let $X \subset \R^d$ be a set of $|X| = n$ points and let $f \colon X \to [k]$ be a labeling induced by $k$ boxes. Given a fixed integer $0 \leq m \leq n$, let $Z \subseteq X$ be a random subset of $X$ of size exactly $m$. Let $Q$ be a random variable denoting the number of points whose label cannot be inferred from $f(Z)$. Then,
    \[
    \mathbb{E}_{Z}[Q] \leq 2d \sum_{j=1}^k \min\left(|f^{-1}(j)|,\frac{n-m}{m+1}\right) \leq 2dk \cdot \frac{n-m}{m+1}.
    \]
\end{lemma}

\begin{proof} 
For each label $y \in [k]$, define $X_y$ and $Z_y$ to be the points with label $y$ in $X$ and $Z$, respectively. Let $n_y = |X_y|$ be the number of points in $X_y$. Let $Q = \sum_y Q_y$ where $Q_y$ is the number of points in $X_y$ whose label cannot be inferred from the labels $f(Z)$, where
\[
    \textrm{the label }f(x) = y \text{ can be inferred from } f(Z) ~\Longleftrightarrow~ x \in \mathsf{bbox}\big(Z_y\big) = \prod_{i=1}^d [a_{y,i}, b_{y,i}].
\]
For each dimension $i \in [d]$, we also count the number of points in $X_y$ that land to the left and right of $\mathsf{bbox}(Z_y)$ in the $i$th dimension:
\[L_{y,i} = \# \big\{x \in X_y : x_i < a_{y,i}\big\} \qquad\textrm{and}\qquad R_{y,i} = \# \big\{x \in X_y : x_i > b_{y,i}\big\}.\]
When $Z_y = \emptyset$, we use the convention $L_{y,i} = R_{y,i} = n_y$. We can upper bound $Q_y$ by the sum
\[
    Q_y \leq \sum_{i=1}^d L_{y,i} + R_{y,i}.
\]
We now bound $\mathbb{E}_{Z}[L_{y,i}]$. The bound on $\mathbb{E}_{Z}[R_{y,i}]$ is the same and has an analogous argument.

For each dimension $i \in [d]$, let $\mathsf{rank}_{y,i}(x)$ denote the ordering of the points $x \in X_y$ with respect to the $i$th coordinate, breaking ties arbitrarily. For each integer $\ell \geq 0$, define the $\ell$ leftmost set of points under this ordering to be $X_{y,i}^{\leq \ell} = \{x \in X_y : \mathsf{rank}_{y,i}(x) \leq \ell\}$. Then:
\[L_{y,i} \geq \ell \qquad \implies \qquad Z \cap X_{y,i}^{\leq \ell} = \emptyset.\]
To see this, suppose that $L_{y,i} \geq \ell$. Every point $z \in Z_y$ must strictly dominate a set of $L_{y,i}$ points with respect to $\mathsf{rank}_{y,i}$, which means that $\mathsf{rank}_{y,i}(z) > \ell$. As $Z$ is chosen uniformly at random,
\[\hspace{2em}\Pr_Z\bigg[L_{y,i} \geq \ell\bigg] \leq \Pr_Z \bigg[Z \cap X_{y,i}^{\leq \ell} = \emptyset\bigg] \leq \frac{\binom{n - \ell}{m}}{\binom{n}{m}},\qquad \textrm{for all } 1 \leq \ell \leq n_y.\]
The last inequality holds because $Z$ must miss a set of size $\ell$. Using a layer-cake representation and applying the hockey-stick identity, we obtain
\begin{align*}
    \mathbb{E}_{Z}\big[L_{y,i}\big] &\leq \sum_{\ell=1}^{n - m} \Pr_Z\big[L_{y,i} \geq \ell \big]
    \leq \sum_{\ell=1}^{n - m} \frac{\binom{n - \ell}{m}}{\binom{n}{m}}
    = \sum_{t=m}^{n - 1} \frac{\binom{t}{m}}{\binom{n}{m}}
    = \frac{\binom{n}{m + 1}}{\binom{n}{m}} = \frac{n - m}{m + 1}.
\end{align*}
Putting everything together observing that $Q_y \leq n_y$ always holds, we have
\[\mathbb{E}_Z\big[Q\big] = \sum_{y \in [k]} \mathbb{E}_Z\big[Q_y\big] \leq 2d \sum_{y \in [k]} \min\left\{n_y, \frac{n-m}{m+1}\right\},\]
which completes the proof.
\end{proof}

\begin{proofof}{\Cref{thm:r-round}}
The algorithm, defined as follows, can be seen as a batched-variant of CAL. Each round, instead of picking a single point from the disagreement region, randomly selects a subset of points. To formally define the algorithm, pick a permutation $\pi \colon [n] \to X$, then batch the points into $r$ sets. In the $t$-th round where $1 \leq t \leq r$, query all points in the $t$-th batch whose labels cannot already be inferred from previous queries. We will fix a sequence of $r$ batch sizes so that the expected query complexity over $\pi$ is minimized. Towards this, given an arbitrary schedule
\[
0 = s_0 < s_1 < \cdots < s_r = n,
\]
the $t$-th batch will be $(\pi(i) \colon s_{t-1} < i \leq s_t)$. We first obtain an expression for the expected query complexity of a general schedule using \Cref{lem:expected-coverage}. For $1 \leq t \leq r$, let $Q_t$ be the random variable denoting the number of queries made in the $t$-th round (over the randomness of $\pi$). For $t \geq 2$, conditioned on $Y \subseteq X$ being the first $|Y| = s_{t}$ points in the ordering induced by $\pi$, the first $s_{t-1}$ such points form a uniformly random subset of $Y$ of size $s_{t-1}$. Therefore, to bound $\mathbb{E}_{\pi}[Q_t]$ we can simply apply \Cref{lem:expected-coverage} with $s_t,s_{t-1}$ playing the roles of $n,m$ respectively. We obtain
\begin{align} \label{eq:schedule-bound}
    \mathbb{E}_{\pi}[Q] = s_1 + \sum_{t=2}^r\mathbb{E}_\pi[Q_t] \leq s_1 + 2dk\sum_{t=2}^r \frac{s_t - s_{t-1}}{s_{t-1} + 1} \leq s_1 + 2dk\sum_{t=2}^r \frac{s_t}{s_{t-1}+1}
\end{align}
To choose a schedule that optimizes the above expected query complexity, for each $1 \leq t \leq r$, set 
\begin{align} \label{eq:optimal-schedule}
    s_t = \left\lfloor 2dk \cdot \left(\frac{n}{2dk}\right)^{t/r} \right\rfloor.
\end{align}
Substituting this into \Cref{eq:schedule-bound} yields
\[
\mathbb{E}_{\pi}[Q] \leq 2dk \cdot \left(\frac{n}{2dk}\right)^{1/r} + 2dk \sum_{t=2}^r \left(\frac{n}{2dk}\right)^{1/r} = 2dk r \left(\frac{n}{2dk}\right)^{1/r}
\]
as claimed.

\end{proofof}

\begin{remark} The optimal $O(kd \log(n/dk))$ query complexity can be obtained by setting the round complexity to only $r = \log_2(n/2dk)$ in \Cref{thm:r-round}. From \Cref{eq:optimal-schedule} in the proof, we see that the $t$-th round considers points in a batch of size 
\[
s_t - s_{t-1} \approx 2dk \cdot 2^{t-1} 
\]
and the number of queries made in that batch is expected to be 
\[
\lessapprox 2dk \cdot \frac{s_t}{s_{t-1}} \approx 2kd.
\]
\end{remark}
\section{Separation between CAL and Halving++} \label{sec:separation}

This section gives a construction of a disjoint-box problem where CAL uses strictly more queries in expectation than Halving++. We restate the query complexities from \Cref{sec:general-box-partitions,sec:disjoint-box-partitions} here. Treating the number of boxes $k = O(1)$ as a constant, they are
\[\hspace{1em}\textrm{CAL: } O_k\big(d \log (n/d)\big) \textrm{ queries}\hspace{3em} \textrm{Halving++: } O_k\big(d \log(n)/\log(d)\big) \textrm{ queries}.\]
While it seems that Halving++ attains a $\log(d)$-factor saving, these bounds cannot be directly compared because the bound for CAL applies to general boxes, while the one for Halving++ applies to disjoint boxes; we cannot deduce that Halving++ is more label-efficient than CAL. 

Nevertheless, the next result strictly separates these two algorithms. We construct a disjoint-box problem where the upper bound we proved for CAL is tight. In particular, if we let $n = d^2$, then CAL requires $\Omega(d \log d)$ queries. In contrast, for the same problem, the above query complexity of the more careful Halving++ algorithm resolves to $O(d)$ queries.

\begin{proposition}[Query lower bound for CAL]
    Let $k \geq 2$. There is an $n$-point set $X$ in $\mathbb{R}^d$ so that the expected number of queries that CAL requires to learn any $k$-disjoint box partition over $X$ is
    \[\hspace{1em}\Omega(kd \log (n/kd)) \textrm{ queries}.\]
\end{proposition}

\begin{proof}
    We first give a proof when $k = 2$ and $d = 1$. Let $X = [n]$. Unbeknownst to the learner, let the labeling $f(x) \equiv y$ be a constant map that assigns every point to the same class. This is a valid two-disjoint box partition, where one of the boxes is empty.

    Let's see how CAL works; it is also described in the proof of \Cref{thm:overlap}. Say CAL makes a sequence of queries $(q_t)_t$. At the end of round $t$, it can infer the labels of points only in the set
    \[X \cap \mathsf{bbox}(q_1,\ldots, q_t).\]
    This is because any point $x \in X \setminus \mathsf{bbox}(q_1,\ldots, q_t)$ can be split off from the queried points via a threshold; there is a valid two-disjoint box partition that is consistent with the observed labels that assigns $x$ a different label. And so, in the next round, CAL chooses $q_{t+1}$ uniformly at random from the points not covered by the previous bounding box, and CAL terminates only when the bounding box covers every point in $X$. It follows that the query complexity $Q_\mathrm{CAL}$ of CAL for this problem instance is simply the amount of time it takes for the bounding box over the queries to cover $X$. 
    
    We can compute the expectation $\mathbb{E}[Q_\mathrm{CAL}]$ as follows. Let $\pi : [n] \to X$ be a random ordering of the points in $X$, where CAL iterates through $\pi(s)$ and queries it if it cannot yet infer its label. Notice that CAL will query a point $i \in X$ if it has not previously a point to the left:
    \[\forall j < i,\quad \pi^{-1}(i) < \pi^{-1}(j) \qquad \implies \qquad \textrm{ CAL queries $i$}.\]
    Since the ordering is random, we have
    \[\mathbb{E}[Q_\mathrm{CAL}] \geq \sum_{i = 1}^n \Pr(\textrm{CAL queries $i$}) \geq \sum_{i=1}^n \frac{1}{i} = H(n),\]
    where $H(n) \geq \log(n)$ is the $n$th harmonic number.

    In higher dimensions, we can improve the bound by making a copy for each dimension. Define
    \[X = \bigcup_{i=1}^d \big\{\alpha e_i : \alpha \in [n/d]\big\}.\]
    When $k =2$, let $f(x) \equiv y$ be constant. As in the 1-dimensional case, CAL can only infer the labels of points contained in the bounding box of its previous queries: there is always an axis-aligned halfspace splitting $\mathsf{bbox}(q_1,\ldots, q_t)$ from any point from $X$ not contained in it. Notice that if CAL queries a point $\alpha e_i$ lying along the $i$th dimension, the bounding box only grows in the $i$th dimension, and so we may treat each dimension independently. This gives a lower bound 
    \[\mathbb{E}[Q_\mathrm{CAL}] \geq d \log(n/d).\]

    Finally, with general $k \geq 2$, we can make $(k-1)$ translated copies of this construction, where each copy is assigned a separate label. The same argument yields the claimed lower bound
    \[\mathbb{E}[Q_\mathrm{CAL}] \geq (k-1)d \log (n/kd).\]
\end{proof}

\end{document}